\pdfoutput=1
\documentclass[iop, apjl, twocolappendix, numberedappendix]{emulateapj}
\usepackage[T1]{fontenc}
\usepackage{float}
\usepackage{graphicx}
\usepackage{amsmath}
\usepackage{amssymb}
\usepackage{txfonts}
\usepackage{ulem}
\usepackage{xspace}
\usepackage[colorlinks=true,urlcolor=blue,linkcolor=blue,citecolor=blue]{hyperref}
\usepackage[utf8]{inputenc}
\newcommand{\fhat}[1]{\expandafter\hat#1}
\newcommand{\fbar}[1]{\expandafter\bar#1}
\newcommand\fnurl[2]{%
  \href{#2}{#1}\footnote{\url{#2}}%
}

\def\mpch{h^{-1}{\rm Mpc}}
\def\msunh{h^{-1}{\rm M_{\odot}}}
\shorttitle{Splashback radius of Planck-SZ clusters}
\shortauthors{Zürcher, More}

\begin{document}

\title{The Splashback Radius of Planck-SZ clusters}

\thanks{This work was submitted and defended as a Masters thesis by Dominik
Zuercher at the DPHYS in ETH Zurich in June 2018.}

\author{
Dominik Zürcher \altaffilmark{1, 2, 4}
Surhud More \altaffilmark{3, 1}, 
}
\affil{
$^{1}$Kavli Institute for the Physics and Mathematics of the Universe (Kavli IPMU, WPI), 5-1-5 Kashiwanoha, Kashiwa, Chiba 277-8583, Japan \\ 
$^{2}$Department of Physics, ETH Zurich, Wolfgang Pauli Strasse 27, 8093 Zurich, Switzerland\\
$^{3}$The Inter-University Centre for Astronomy and Astrophysics, Post Bag 4, Ganeshkhind, Pune 411007, India\\
$^{4}$ dominikz@phys.ethz.ch
}

\begin{abstract}
We present evidence for the existence of the splashback radius in
galaxy clusters selected using the Sunyaev-Zeldovich effect.  We show
that the deprojected cross-correlation of galaxy clusters found in the
Planck survey with galaxies detected photometrically in the Pan-STARRS
survey, shows a sharp steepening feature (a logarithmic slope steeper
than $-3$), which we associate with the splashback radius. We infer
the three-dimensional splashback radius for the SZ cluster sample to
be $r_{\rm sp}=1.85_{-0.30}^{+0.26}$ $\mpch$, where the cluster sample
has an average halo mass $M_{\rm 500c}=3.0\times10^{14}\msunh$ at an
average redshift of $z=0.18$. The inferred value of the splashback
radius appears consistent with the expected location for dark matter halos
in the standard cold dark matter paradigm. However, given the limited
precision of our measurements, we cannot conclusively rule out the
smaller splashback radius measured so far in the literature for
optically selected galaxy clusters. We show that the splashback radius
does not depend upon the galaxy magnitude for galaxies fainter than
$M_i-5\log h=-19.44$, and is present at a consistent location in
galaxy populations divided by color. The presence of the splashback
radius in the star-forming galaxy population could potentially be used to put
lower limits on the quenching timescales for galaxies. We can marginally rule
out the contamination of the star-forming galaxy sample by quenched galaxies,
but the results would need further verification with deeper datasets.

\end{abstract}


\section{Introduction}
\label{sec:Introduction}

The density distribution of matter within dark matter halos shapes
the potential well in which galaxies form and grow. Therefore, the
structure of these dark matter halos has been extensively studied
both theoretically as well as in numerical simulations \citep[see
e.g.,][]{gunn1972infall, fillmore1984self, bertschinger1985self,
navarro1997universal, moore1999dark}. Studies with numerical
simulations show that the density profiles of dark matter halos
within their virial radii are roughly self-similar and follow the
Navarro-Frenk-White (NFW) profile \citep{navarro1997universal},
which asymptotes to a slope of $-1$ in the inner regions and $-3$ at
large radii. There has been intense debate in the literature about
the exact form of the density profile 
\citep[e.g.,][]{navarro2004inner}, the value of the asymptotic inner slope,
as well as the outskirts and boundaries of dark matter halos
\citep{cuesta2008virialized, more2011overdensity, diemer2013pseudo}. 

The recent study of \citet{diemer2014dependence} has sparked a renewed
interest in understanding the structure of dark matter halos on scales
beyond the typical virial radii.  \citet{diemer2014dependence}
investigated the outskirts of dark matter halos in numerical
simulations and found the existence of a physical feature, namely a
sharp steepening in the density distributions of dark matter halos,
which is not captured by commonly used functional forms such as the
NFW profile. They showed that even for halos of the same mass, the
position of the feature changes depending upon the mass accretion rate
of the halos. A simple theoretical toy model to explain this feature
was presented by \citet{adhikari2014splashback}. They showed that the
feature observed by \citet{diemer2014dependence} results from the
piling up of recently accreted dark matter particles at the apocenters
of their orbits, and its location corresponds to the last density
caustic in the self-similar models of secondary infall
\citep{fillmore1984self, bertschinger1985self, lithwick2011self}.
They coined the term ``splashback radius`` for this feature. Their toy
model also naturally explains the accretion rate dependence -- faster
accreting halos have smaller splashback radii. Subsequently,
\citet{more2015splashback} suggested the use of the splashback radius
as a natural boundary for dark matter halos and explored its
consequences for the inferred boundaries and growth rates of the
halos. Depending on the accretion rate of the halo, the splashback
radius can lie well beyond the commonly used virial radius
\citep{diemer2014dependence, more2015splashback}.

The interpretation of the accretion rate dependence is
straightforward. Due to the continuous change of the gravitational
potential of the halo, depending on its accretion rate, the kinetic
energy of a recently accreted dark matter particle, gained during
its infall onto the cluster, does not suffice to climb the deepened
potential well completely again, but instead it ``splashes back`` at
a distance that depends on the recent deepening of the potential
well.

Given the mass of the halo, the location of the splashback radius
constitutes a direct probe of the halo accretion rate. Motivated by
these studies, \citet{more2016detection} attempted to detect this
feature in observations. Using the optically selected Sloan Digital
Sky Survey RedMaPPer galaxy cluster catalog
\citep{rykoff2014redmapper}, and by cross-correlating it with the
SDSS photometric galaxy sample, \citet{more2016detection} found evidence
for the steepening of the dark matter density profile, and therefore
the splashback radius of this sample of galaxy clusters. This was 
corroborated by including further models for mis-centering by
\citet{baxter2017halo} and in the Dark Energy Survey data by
\citet{chang2017splashback} using optically selected clusters. Somewhat surprisingly,
\citet{more2016detection} found that the location of the splashback radius was
inconsistent with that expected from numerical simulations of dark matter by
about $20\pm5$\% (see also \citet{baxter2017halo, chang2017splashback}).
Although they
investigated potential systematic issues, they did not have access
to mock cluster catalogs which could mimic the selection effects of
optically-identified clusters. \citet{busch2017assembly} used a
simplified optical cluster selection algorithm on the Millennium
simulation, and pointed out that optical clusters can be heavily
affected by projection issues, and could potentially introduce
systematics in the inference of the splashback radius, as well as
halo assembly bias. The existence of projection effects in the
optical cluster catalog in the context of halo assembly bias was 
also demonstrated by \citet{zu2016level}.

\citet{umetsu2017lensing} as well as \citet{contigiani2018weak}
recently used Xray selected clusters to look for the splashback radius
using the weak lensing signal, however stacking issues and the low
signal-to-noise ratio remains a significant hurdle for both of them.
\citet{chang2017splashback} found evidence for the splashback radius
in the weak lensing signal, but their analysis was again done using
optically selected clusters in the Dark energy survey. Regardless of the
projection issues present in the optical cluster catalog, there is some
inherent circularity present in the logic of using photometric galaxies to
select clusters as over-densities in a given aperture, and then using the same
photometric sample of galaxies to look for the splashback radius. There is a
possibility that the aperture used to select the cluster catalogs could be
imprinted in a non-trivial way on the measured number density profiles of
clusters.

In this work, we move away from the optical cluster selection and
explore the use of SZ selected cluster catalogs. While the SZ
selected clusters can also be susceptible to systematic selection
effects, the scales on which the SZ signal is measured and the
cluster selection is performed is much smaller than the expected
location of the splashback radius (typically $R_{\rm 500c}$). We use
this sample to explore the evidence for the splashback radius in
observations.  Due to the low signal-to-noise ratio of the weak
lensing signal, we perform our analysis using cross-correlation of
galaxies with clusters.  The SDSS sample used by
\citet{more2016detection} was deep enough such that biases in the
location of the splashback radius due to dynamical friction effects
were expected to be small.  Nevertheless, we use galaxy samples, which
are even fainter by $0.5-1$ magnitudes compared to those used by
\citet{more2016detection}.

Given that the splashback radius represents a true halo boundary,
the observations of the splashback radius can be used to study a
variety of galaxy formation questions. Questions regarding the
timescales and the spatial scales within which star forming galaxies
quench after they fall into the cluster potential are of particular
interest to understand the fate of star formation in satellite
galaxies. In particular, if star forming galaxies quench before they
reach the apocenters of their orbit after infall, then they are not
expected to show a splashback feature in their density distribution.
In pursuit of this question, we also explore the dependence of the
cluster-galaxy cross-correlations separately for star forming and
quenched galaxy populations as separated by their color.

This paper is organized as follows. We introduce the observational
data sets we use in Section~\ref{sec:data}, namely the cluster and the
galaxy catalogs. We describe the methods and analysis procedures we
use in Section~\ref{sec:estimators}. We present and discuss our
results in Section~\ref{sec:Results}. Finally, we summarize our
findings in Section~\ref{sec:Conclusions} and discuss possible future
directions. Throughout the paper, we use a flat $\Lambda$CDM cosmology
with $\Omega_{\rm m}=0.27$ and a dimensionless Hubble parameter of $h=0.7$ 
to convert redshifts and angles into
cosmological distances. Also, we denote three-dimensional distances by
$r$ and projected distances by $R$.

\section{Data}
\label{sec:data}

\subsection{Cluster catalog}
\label{sec:clusters}

\begin{figure}
    \includegraphics[scale=0.55]{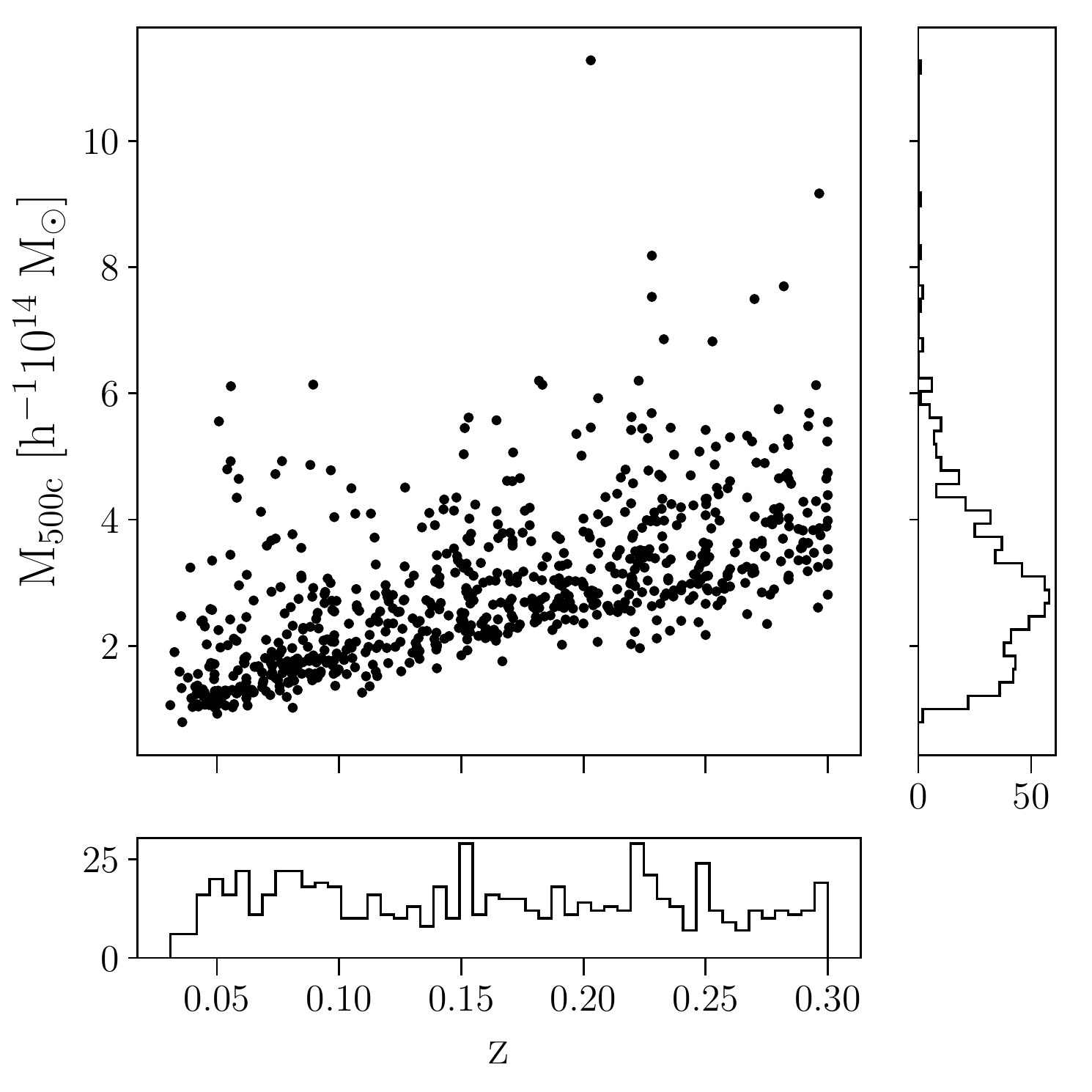}
\caption{The distribution of the masses and redshifts of the galaxy clusters
from the PSZ2 catalog that we use in our analysis. The histograms on the right
hand side and the bottom show the distribution in mass and redshift,
respectively.}
   \label{fig:planck_summary}
\end{figure}

The baryonic component of a galaxy cluster is dominated by the hot,
ionized  intra-cluster medium (ICM), which is gravitationally bound
within the cluster. The cosmic microwave background (CMB) photons
that pass through the cluster inverse Compton scatter off the hot
electrons and gain energy. This effect is known as the thermal
Sunyaev-Zeldovich (SZ) effect
\citep{sunyaev1970small,sunyaev1980velocity}. The effect has a
characteristic frequency dependence and results in an intensity
decrease below 220 GHz and an associated increase at higher
frequencies. The multiple frequency channels on the Planck satellite
allow a detection of galaxy clusters using the SZ effect
\citep{collaboration2016planck}. As part of the 2015 Data Release of
the Planck mission, the second Planck Catalogue of Sunyaev-Zeldovich
Sources (PSZ2) was made available to the community. The PSZ2 catalog
contains detections based on  three different techniques
\citep{ade2016planck}, and the union of these catalogs has in total
1653 galaxy clusters, of which 1203 clusters have been confirmed by
cross-matching to other galaxy clusters from external data sets. 
\begin{table}
    \centering
    \caption{Comparison of the PSZ2 cluster catalog
    against the RedMaPPer cluster catalog used by
\citet{more2016detection}. The values of the mass estimates
$M_{\mathrm{500c}}$, $M_{\mathrm{200m}}$, redshifts $z$ and expected
splashback radii $r^{\mathrm{3D,theo}}_{\mathrm{sp}}$ represent the
catalog averages. The redshifts for both catalogs are given from the
survey. For the RedMaPPer catalog the $M_{\mathrm{200m}}$ mass
estimates were obtained from gravitational lensing and for the PSZ2
catalog the $M_{\mathrm{500c}}$ estimates were calculated from the
survey parameters using the scaling relation between the integrated
Compton Y-parameter $Y_{500c}$ and $M_{500c}$ as found by
\citet{ade2014planck}. The missing mass estimates as well as the
expected values of the splashback radii were calculated using the
Python package COLOSSUS \citep{diemer2017colossus}. The predictions
for the splashback radii $r^{\mathrm{3D,theo}}_{\mathrm{sp}}$ are
given in comoving units.}
    \label{tab:cluster_catalogs} 
    \begin{tabular}{ccc}
    \hline 
    & RedMaPPer & PSZ2 \\
    \hline 
    $z$ & 0.24 & 0.177 \\
    \hline 
    \# objects & 8643 & 596\\
    \hline
    $M_{\mathrm{500c}}$ [$h^{-1}10^{14} $M$_{\sun}$] & 0.9 & 3.0 \\
    \hline
    $M_{\mathrm{200m}}$ [$h^{-1}10^{14} $M$_{\sun}$] & 1.8 & 6.2\\ 
    \hline
    $r^{\mathrm{3D,theo}}_{\mathrm{sp}}$ [$h^{-1}$ Mpc] & 1.37 & 1.89 \\
    \hline
    \end{tabular} 
\end{table}
\begin{table*}
    \centering
    \caption{Summary and comparison of the properties of the galaxy catalogs used in this work as well as the SDSS catalog used in \citet{more2016detection}.}
    \label{tab:galaxy_catalogs}
    \begin{tabular}{ccccc}
    \hline 
    & SDSS & PS 21 & PS 21.5 & PS 22 \\ 
    \hline 
    depth [mag] & 21.00 & 21.00 & 21.50 & 22.00\\ 
    \hline 
    eff. area [deg$^2$] & $\approx$ 10'000 & 21'148 & 20'586 & 15'689\\ 
    \hline 
    \# objects & 57'181'113 & 93'772'329 & 123'188'529 & 105'809'113 \\
    \hline
    objects/deg$^2$ & 5718 & 4434 & 5984 & 6744\\ 
    \hline
    \end{tabular} 
\end{table*}
\begin{table*}
    \centering
    \caption{Summary of the priors used in the MCMC sampling
procedure. The MCMC chains are constrained using flat priors or
normal priors on some of the fitting parameters. This table lists
the ranges of the flat priors and the central positions as well as
the scales of the normal priors ( format: center | scale ),
respectively.}
    \label{tab:priors}
    \begin{tabular}{ccccccccc}
    \hline 
    & $\log_{10}(\rho_{\mathrm{s}})$ & $\log_{10}(\alpha)$ & $\log_{10}(r_{\mathrm{s}})$ & $\rho_0$ & $s_{\mathrm{e}}$ & $\log_{10}(r_{\mathrm{t}})$ & $\log_{10}(\beta)$ & $\log_{10}(\gamma)$ \\ 
    \hline 
    Prior Type & None & Normal & Flat & None & None & Flat & Normal & Normal\\ 
    \hline 
    Prior Range & - & $\log_{10}(0.2)| 1.2$ & [0.1,5.0] & - & - & [0.1,5.0] & $\log_{10}(6.0) | 0.4$ & $\log_{10}(4.0) | 0.4$ \\
    \hline
    \end{tabular} 
\end{table*}

The 1-$\sigma$ errors on the cluster positions are $\sim1.6\arcmin$
and the estimated purity of the catalog has a lower limit of 83\%
\citep{ade2016psclusters}. The integrated Compton Y-parameter
$Y_{500c}$ of each of the clusters are also provided. The mass
estimates $M_{500c}$ provided by the Planck collaboration are based on
the scaling relation between $Y_{500c}$ and $M_{500c}$
\citep{ade2014planck,adam2016planck,collaboration2016planck}. The PSZ2
union cluster catalog is publicly available from the
\fnurl{\textit{Planck Legacy Archive}}{https://pla.esac.esa.int/pla/\#
home}.

We restrict ourselves to the redshift range $0.03 \leq z \leq 0.33$ in order to
have a similar redshift range used in \citet{more2016detection}. The mass and
the redshift distribution of the Planck clusters we use is shown in
Fig.~\ref{fig:planck_summary}. Due to the larger beam size of the Planck
satellite, the cluster positions as reported in the catalog may be
mis-centered from the true centers of the galaxy clusters. We perform
a visual inspection of Pan-STARRS images taken around the detected
galaxy clusters in order to locate the nearest, brightest cluster
galaxy (BCG). We regard the position of the BCG as the true cluster
location, under the assumption that the BCG is located at the true,
gravitational center. Due to this assumption, we additionally 
study the effects of a possible, remaining mis-centering of the 
cluster positions in our model for the two-dimensional correlation 
function (see Appendix~\ref{sec:mis-centering}). 

The final sample that we use consists of 596
galaxy clusters and is about an order of magnitude smaller compared
to the sample used in \citet{more2016detection}. The sample we use
in this paper has an average redshift of 0.177, and an average
cluster mass $M_{500c}$ of about $3.0 \times 10^{14}
h^{-1}$M$_{\sun}$. In Table~\ref{tab:cluster_catalogs} we compare
the main properties of the cluster catalog used in this work to that
used by \citet{more2016detection}. Additionaly, Figure~\ref{fig:planck_summary}
visualizes the distribution of the masses and redshifts of the used clusters.
A map of the sky positions of the clusters
can be found in Figure~\ref{fig:planck_fig} in
Appendix~\ref{sec:figures}. 

\citet{kosyra2015environment} found no evidence for a significant
correlation between the density of Planck detections and the
weighted average noise of all Planck channels at $z<0.5$. Since we
restrict ourselves to $z<0.33$, we utilize the selection mask of the
PSZ2 union catalog in order to construct a random galaxy cluster
catalog, which is roughly one order of magnitude larger than the
original catalog. The redshifts of these random objects are drawn
from the parent cluster catalog in order to match the redshift
distribution of the original cluster catalog. 

\subsection{Galaxy catalog}
\label{sec:galaxies}
The Panoramic Survey Telescope and Rapid Response System
(Pan-STARRS) is a wide-field astronomical imaging and data
processing facility operated by the University of Hawaii's Institute
for Astronomy \citep{kaiser2002pan,kaiser2010pan}. We use data from
the 3$\pi$ Steradian Survey carried out with this facility, which
was released as part of Data Release 1 (DR1). The survey covers the
entire sky north of $\delta=-31\degr$ (in ICRS coordinates) in five
broadband filters ($g_{\mathrm{P1}}, r_{\mathrm{P1}},
i_{\mathrm{P1}}, z_{\mathrm{P1}}, y_{\mathrm{P1}}$) with multiple
pointings. The mean 5$\sigma$ point source limiting sensitivities
amount to (23.3, 23.2, 23.1, 22.3, 21.4) magnitudes for the
individual bands, respectively. 

For the visual inspection and centering of the clusters, we use the
Pan-STARRS $gri$ \textit{stack} images around each cluster position.
The galaxy catalog is obtained from the  \textit{StackObjectThin}
table, which is publicly available on the \fnurl{\textit{Barbara A.
Mikulski Archive for Space Telescopes}
(MAST)}{http://archive.stsci.edu/}. To select only objects detected
with acceptable precision we restrict our search to those objects
which have been flagged as \textit{BestDetection}s. We further
restrict the catalog to objects flagged as
\textit{PrimaryDetection}s in order to select unique objects. This
is necessary since the survey is divided into overlapping
\textit{projectioncells} and \textit{skycells}, which causes some objects
to be listed multiple times.

The magnitudes of the selected objects are then corrected for the
extinction caused by dust present in the Milky Way. This is done
using the \fnurl{mwdust}{https://github.com/jobovy/mwdust} Python
module provided by \citet{bovy2016galactic}. The extinction
correction is performed using a dust map combining the measurements
of \citet{marshall2006modelling}, \citet{green2015three} and
\citet{drimmel2003three}. Only objects with an extinction corrected
$i_{\mathrm{P1}}$ band Kron magnitude brighter than 22.0 are
selected from the catalog. Starting from this catalog we construct
three different sub-catalogs corresponding to survey depths of 21.0,
21.5 and 22.0 magnitudes, and we name these catalogs PS 21, PS 21.5
and PS 22, respectively.

As mentioned in the description of the Pan-STARRS survey by
\citet{chambers2016pan} there is a significant variation in the
depth of the 3$\pi$ Steradian survey even on small scales. In order
to avoid choosing objects in shallow regions of the survey the
maximum observed Kron magnitude in the $i_{\mathrm{P1}}$ band in
each \textit{skycell} is recorded and only objects in skycells with
a maximum observed Kron magnitude of 21.0, 21.5 and 22.0 or brighter
are selected depending on the corresponding catalog. Since most of
the shallow regions lie in the galactic plane, we mask out the
region at low galactic latitudes $|b|<20\degr$. The resultant
HEALPix maps showing the excluded areas on the sky can be found in
Figure~\ref{fig:heal_map}. We further disregard objects in bad
pixel regions as indicated by the $i_{\mathrm{P1}}$ band
\textit{stack.mask} images.

At this point of the analysis, our object catalogs contain both
galaxies and stars. The 3$\pi$ Steradian Survey provides both the
Kron and PSF model based magnitudes for each object. These
magnitudes are expected to be similar for stars while the Kron
magnitudes are brighter for galaxies. Therefore, we flag all objects
with a value of $i_{\mathrm{P1,PSF}} - i_{\mathrm{P1,Kron}}< 0.05$
as stars \citep{farrow2013pan}. Despite this cut, bright, close-by
stars at magnitudes brighter than 13.5 tend to be classified as
extended objects \citep{chambers2016pan}. To avoid contamination of
the galaxy catalog due to such bright stars, we further remove all
objects with $i_{\mathrm{P1,PSF}} < 15.0$. Since there are very few
galaxies at such low magnitudes this does not introduce a selection
bias. We list the main characteristics of the galaxy catalogs we
have used in Table~\ref{tab:galaxy_catalogs}.

\section{Methods}
\label{sec:estimators}
The methodology we adopt for locating the splashback radius closely
follows that of \citet{more2016detection}. We perform a
cross-correlation between SZ selected galaxy clusters with photometric
galaxies in order to assess the existence and location of the
splashback radius. We use the Davis-Peebles estimator
\citep{davis1983survey} to compute the cross-correlation between our
galaxy clusters and galaxies. This estimator can be written as
\begin{equation}
\xi_{\rm 2D}(R) = \frac{\rm D_1D_2-R_1D_2}{\rm R_1D_2}\,
\end{equation}
where D$_1$D$_2$ and R$_1$D$_2$ are the normalized numbers of
cluster-galaxy pairs and cluster randoms-galaxy pairs at a given
comoving projected separation $R$. The subtraction of the signal
around random cluster positions gets rid of the uncorrelated pairs and allows
us to estimate the projected cross-correlation. The uncertainty in the galaxy
distribution masks prevents us from using the Landy-Szalay estimator \citep{landy1993bias}. 

Given the flux limited galaxy catalog that we use, we expect to
observe more correlated galaxies in galaxy clusters that lie closer
to us, but with much fainter absolute magnitudes. To avoid
such biases with redshift of the clusters we restrict ourselves to
galaxies with absolute magnitudes brighter than a certain magnitude
limit, which depends on the depth of the used galaxy catalog. We
make the assumption that the galaxies reside at the redshift of the
cluster in question. The magnitude limits we use are -19.44, -18.94 and -18.44
and name the corresponding catalogs PS 21, PS 21.5 and PS 22, respectively.

We use the functional form of \citet{diemer2014dependence} in order
to model our two-dimensional correlation function measurements. This
functional form consists of an inner Einasto profile and an outer
power law profile connected by a smooth transition
\begin{align}
\xi_{\rm 3D}(r) &= \rho_{\mathrm{in}}(r)f_{\mathrm{trans}}(r) + \rho_{\mathrm{out}}(r)\,, \label{eq:model} \\
\rho_{\mathrm{in}}(r)&=\rho_{\mathrm{s}}\exp\left( -\frac{2}{\alpha}\left[ \left( \frac{r}{r_{\mathrm{s}}}\right)^{\alpha}-1 \right] \right)\,,\\
\rho_{\mathrm{out}}(r)&=\rho_{\mathrm{0}}\left( \frac{r}{r_{\mathrm{out}}}\right)^{-s_{\mathrm{e}}} \,,\\
f_{\mathrm{trans}}(r)&=\left( 1+\left( \frac{r}{r_{\mathrm{t}}} \right)^{\beta} \right)^{-\gamma/ \beta}\,,
\label{eq:dk14}
\end{align}
where $r$ indicates the three-dimensional radial distance from the
halo center \citep{diemer2014dependence}. We model the
two-dimensional correlation function, $\xi_{\rm 2D}$ as an integral
over the three-dimensional correlation function
\begin{equation}
\xi_{\rm 2D}(R)=\frac{1}{R_{\rm max}}\int^{R_{\mathrm{max}}}_0 \xi_{\rm 3D}(\sqrt{R^2+x^2}) dx \,
\label{eq:surface}
\end{equation}
where we adopt $R_{\mathrm{max}}=40$ $h^{-1}$Mpc for the maximum projection
length. Variations of this length do not change the location of the
splashback radius appreciably as tested in
\citet{more2016detection}. The functional form adopted in
Equation~\ref{eq:surface} has nine model parameters,
$\rho_{\mathrm{s}}, \alpha, r_{\mathrm{s}}, r_{\mathrm{out}},
\rho_{\mathrm{0}}, s_{\mathrm{e}}, r_{\mathrm{t}}, \beta$ and
$\gamma$. Given the perfect degeneracy between the parameters
$r_{\mathrm{out}}$ and $\rho_{\rm o}$, we fix $r_{\mathrm{out}}=1.5$ $
h^{-1}$Mpc. 

Therefore the model is described by eight model parameters.
We infer the posterior distributions of those parameters by fitting the model for the two-dimensional correlation
function to the measured two-dimensional correlation signal. 
We use the affine invariant Markov Chain Monte Carlo sampler of \citet{goodman2010ensemble} as
implemented in the parallel python package {\it emcee} by
\citet{foreman2013emcee}. We adopt priors similar to
\citet{more2016detection} on some of our parameters based on the
expectations of their values from numerical simulations, but double
the scales of the normal priors compared to their work (see
Table~\ref{tab:priors}). The central value for the prior on $\alpha$
is deduced from mass estimates \citep{gao2008redshift}, whereas the
central values on the priors of $\beta$ and $\gamma$ were
recommended by \citet{diemer2014dependence}.

The splashback radius for halos on galaxy cluster scales is consistent
with the location of the steepest logarithmic slope of the density
profile. We estimate the steepest slope of both the two-dimensional
and the three-dimensional cross-correlation function. The
locations of the steepest slope in the two-dimensional and the
three-dimensional case are expected to be different by
about 20\% for typical cluster halo parameters
\citep{diemer2014dependence, more2016detection}.

Mis-centering of the central cluster positions can affect the
small-scale correlation function on scales smaller than the typical
mis-centering distance.  Although, the splashback radius is located at
much larger scales, the change of the inner part of the correlation
function can alter the model fit significantly.  Mis-centering effects
are not expected to change the location of the splashback feature but
they can in principle decrease the significance of the evidence for
the splashback feature \citep{baxter2017halo}.  We discuss the
modelling of mis-centering and its effects on the inferred splashback
radii in Appendix~\ref{sec:mis-centering}.

\subsection{Separation of red and blue galaxies}
\label{sec:Color}

We are also interested in measuring the cluster-galaxy
cross-correlations for the blue and the red galaxy samples
separately. We use a $g_{\mathrm{P1}}-r_{\mathrm{P1}}$ color cut
which varies with the redshift of the clusters in order to account
for the k-corrections, which cannot be computed individually for
each galaxy.

In order to compute the color cut to be used, we first match spectroscopic
galaxies in the Sloan Digital Sky Survey (DR 8) to their
Pan-STARRS photometry. We bin these galaxies in narrow redshift bins and
produce a histogram of the $g_{\mathrm{P1}}-r_{\mathrm{P1}}$ colors of the
galaxies in each bin. Due to the presence of the $4000\AA$-break present in the
quenched galaxy population, the two galaxy populations separate in such histograms
into two populations and we fit a double Gaussian distribution to it.
Based on the fitted distribution we use a cut in color to exclude a
contamination to the star-forming galaxy population with a confidence
of 3$\sigma$. We repeat this procedure for each redshift bin, to obtain a redshift
dependent color cut that separates the red from the blue population minimizing
the contamination of the blue population by red galaxies. Although being much
simpler and faster than calculating the individual k-corrections for each
galaxy, this procedure has its own shortcomings as discussed in
Appendix~\ref{sec:Errors}, in particular due to the photometric errors of red
galaxies. We would therefore exercise appropriate caution while interpreting
the results obtained by dividing galaxies into color bins.

In our analysis, the same galaxy may be considered red or blue depending upon
the cluster redshift under consideration. Our method avoids the use of
uncertain photometric redshifts to derive k-corrections
\citep[cf.][]{baxter2017halo}. We will study the cross-correlations to derive
the splashback radii for these galaxy populations separately. The caveats in
the interpretation of these results due to photometric errors are discussed in the
Appendix~\ref{sec:Errors}.

\section{Results}
\label{sec:Results}
\subsection{Cluster-galaxy cross-correlations}
\begin{figure}
    \includegraphics[scale=0.65]{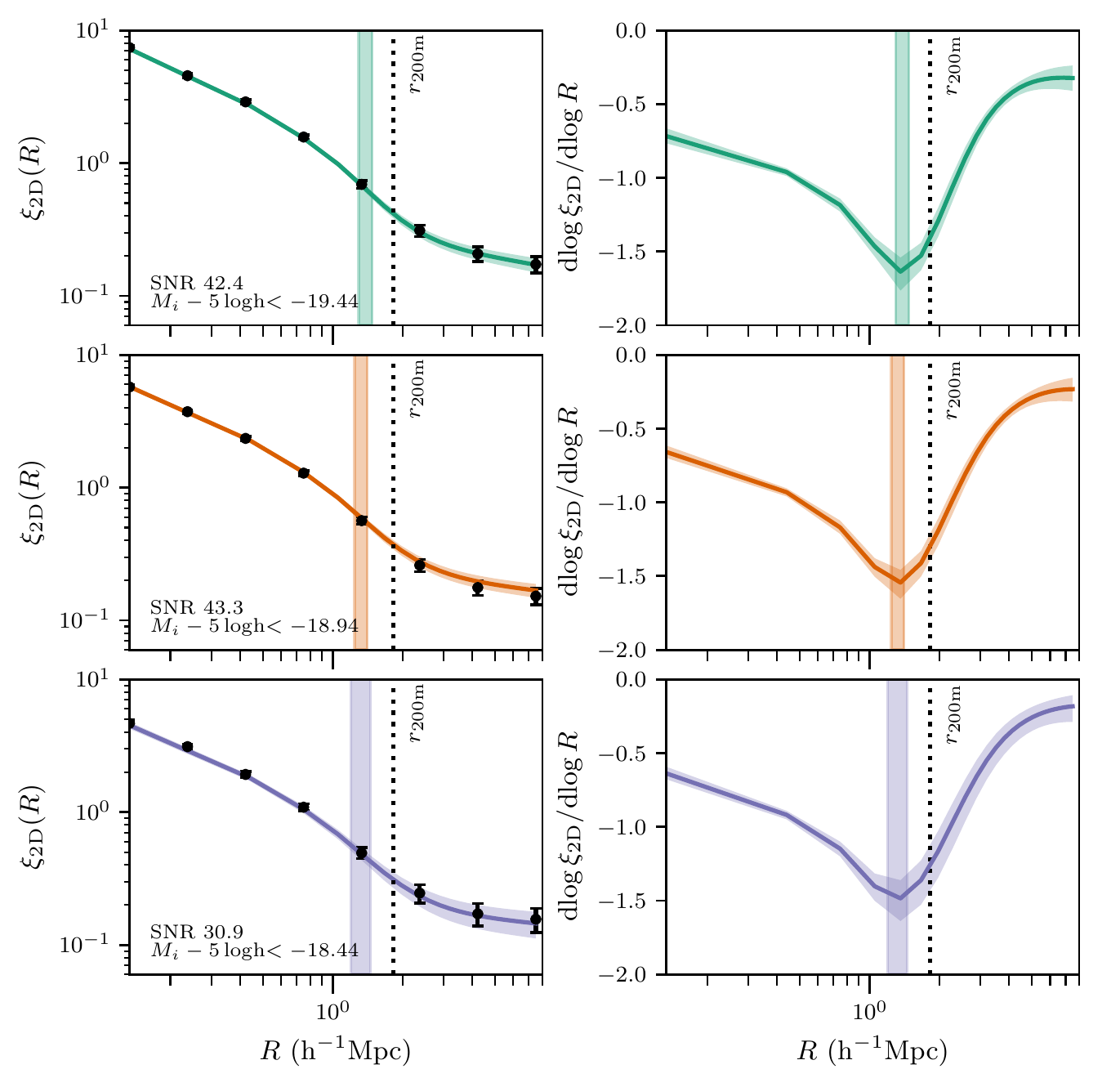}
\caption{The estimates of the two-dimensional cross-correlation
signals (black dots) are shown in the left column. The colored
curves show the two-dimensional model fits of the functional
form in Equation~\ref{eq:surface}. The magnitude limit applied to the galaxy
catalog is indicated in each row along with the signal-to-noise ratio (SNR). 
The vertical, shaded regions indicate the estimates of the locations of 
steepest slope of the profiles as estimated from the corresponding minima of the
derivative profiles, which are shown in the right column. For
comparison, the location of the $r_{\mathrm{200m}}$ radius as
calculated from the average cluster sample properties is indicated
by the black, dotted lines.}
   \label{fig:2D_graphs}
\end{figure}

The measurements of the two-dimensional cluster-galaxy
cross-correlations are shown as black points with error bars in the
left column of Figure~\ref{fig:2D_graphs}. The different rows
correspond to the three different absolute magnitude limits that we
have used to select all the galaxies when calculating the
cross-correlations. The cross-correlation signal is clearly detected
in all three measurements and the corresponding signal-to-noise
ratios of the measurements are listed in Table~\ref{tab:snr}.

We fit these measurements with our parametric model in Equation~\ref{eq:surface}
and compute the posterior distribution of the model parameters 
as described in Section~\ref{sec:estimators}. The
median of the MCMC fit is indicated by the central solid line, while
the shaded area marks the 68\% credible interval for the fit. The
median values of the posterior distributions of the parameters along
with their 68\% confidence intervals are listed in
Table~\ref{tab:fit_parameters} along with the corresponding reduced
$\chi^2$ values for the best fit. The two-dimensional posterior
distributions for each pair of parameters are shown in
Figures~\ref{fig:corner_21}, \ref{fig:corner_21.5} and
\ref{fig:corner_22} for the three absolute magnitude limits,
respectively.

\begin{figure*}
    \centering
    \includegraphics[scale=0.85]{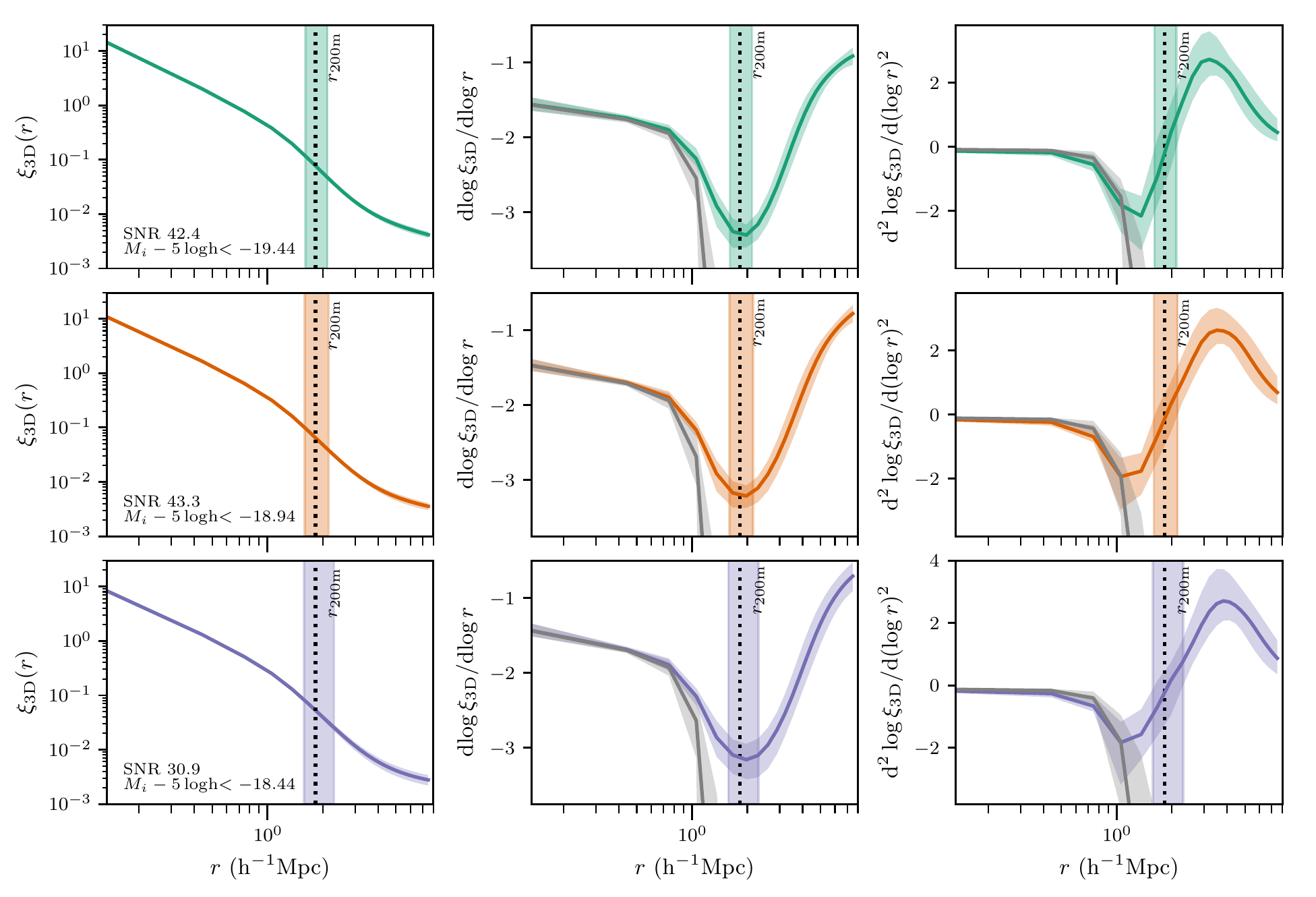}
\caption{The estimates of the three-dimensional cross-correlation
signals are shown in the leftmost column. The colored curves show the
model fits of the functional form in Equation~\ref{eq:model}. The magnitude
limit applied on the galaxy catalog is indicated in each row along with 
the signal-to-noise ratio (SNR). The vertical, shaded regions indicate the 
estimates of the splashback radii as estimated from the corresponding minima 
of the derivative profiles, which are shown in the column in the middle. The
rightmost column shows the second derivative profiles in matching color.
The grey curves show the corresponding, estimated first and second derivatives of the 
one-halo term (namely $\rho_{\mathrm{in}}(r)f_{\mathrm{trans}}(r)$ in Equation~\ref{eq:dk14}).
For comparison, the location of the $r_{\mathrm{200m}}$ radius as calculated 
from the average cluster sample properties is indicated by the black, dotted
lines.}
   \label{fig:3D_graphs} 
\end{figure*}

\begin{table}
    \centering
    \caption{We list the signal to noise ratios (SNRs) of our
different estimates of the two-dimensional cross-correlation
obtained by using the different galaxy samples. For comparison: The SNR
achieved by \citet{more2016detection} is 263. It is much higher
due to size of the cluster sample used in their study.}
    \label{tab:snr}
    \begin{tabular}{cccc}
    \hline
    gal cat & PS 21 & PS 21.5 & PS 22 \\ 
    \hline
    \hline
   SNR & 42.4 & 43.3 & 30.9 \\
    \hline
    \end{tabular} 
\end{table}

\begin{table*}
    \centering
    \caption{The table presents the posterior distribution of the parameters obtained from fitting the cross-correlation of the galaxy cluster sample with the different galaxy catalogs. For each parameter estimate the median as well as the 16\% and 84\% quantiles of the posterior distribution are given. We also list the estimated locations of the steepening feature in the two-dimensional cross-correlation signal ($R_{\mathrm{sp}}^{\mathrm{2D}}$), as well as the three-dimensional splashback radius ($r_{\mathrm{sp}}^{\mathrm{3D}}$). In the last column the minimum, reduced $\chi^2$ value of the model fit is indicated. The two-dimensional posterior distributions of the fitting parameters are displayed in 
Figure~\ref{fig:corner_21} up to Figure~\ref{fig:cornerblue}.}
    \label{tab:fit_parameters}
    \begin{tabular}{cccccccccccccc}
    \hline 
gal cat & $\log_{10}(\rho_{\mathrm{s}})$ & $\log_{10}(\alpha)$ & $\log_{10}(r_{\mathrm{s}})$ & $\rho_{\mathrm{0}}$ & $s_{\mathrm{e}}$ & $\log_{10}(r_{\mathrm{t}})$ & $\log_{10}(\beta)$ & $\log_{10}(\gamma)$ & $R_{\mathrm{sp}}^{\mathrm{2D}}$ & $r_{\mathrm{sp}}^{\mathrm{3D}}$ & $\chi^2/\nu$ \\
\hline 
\hline 
PS 21 & $-2.93_{-0.50}^{+0.44}$ & $-1.06_{-0.18}^{+0.14}$ & $0.35_{-0.24}^{+0.27}$ & $0.000179_{-0.000086}^{+0.000075}$ & $0.78_{-0.19}^{+0.24}$ & $0.123_{-0.122}^{+0.051}$ & $0.74_{-0.31}^{+0.21}$ & $0.35_{-0.21}^{+0.13}$ &$1.384_{-0.096}^{+0.088}$&$1.86_{-0.26}^{+0.25}$& $1.524$ \\
\hline
PS 21.5 & $-2.93_{-0.45}^{+0.45}$ & $-0.95_{-0.16}^{+0.13}$ & $0.32_{-0.25}^{+0.25}$ & $0.000103_{-0.000065}^{+0.000047}$ & $0.56_{-0.22}^{+0.25}$ & $0.095_{-0.115}^{+0.045}$ & $0.74_{-0.32}^{+0.24}$ & $0.27_{-0.19}^{+0.12}$ &$1.323_{-0.086}^{+0.080}$&$1.85_{-0.30}^{+0.26}$& $0.285$ \\
\hline
PS 22 & $-2.95_{-0.44}^{+0.49}$ & $-0.91_{-0.17}^{+0.14}$ & $0.28_{-0.27}^{+0.25}$ & $0.000062_{-0.000051}^{+0.000028}$ & $0.41_{-0.28}^{+0.25}$ & $0.094_{-0.190}^{+0.060}$ & $0.76_{-0.36}^{+0.29}$ & $0.24_{-0.27}^{+0.14}$ &$1.31_{-0.14}^{+0.11}$&$1.90_{-0.40}^{+0.32} $&$0.211$ \\
\hline
PS 21.5 (R) & $-0.69_{-0.54}^{+0.55}$ & $-0.99_{-0.20}^{+0.17}$ & $0.24_{-0.31}^{+0.30}$ & $0.0092_{-0.0065}^{+0.0046}$ & $0.63_{-0.27}^{+0.30}$ & $0.157_{-0.171}^{+0.091}$ & $0.54_{-0.20}^{+0.16}$ & $0.36_{-0.26}^{+0.17}$ & $1.437_{-0.089}^{+0.099}$ & $2.13_{-0.21}^{+0.22}$ & $0.502$ \\
\hline
PS 21.5 (B) & $-1.00_{-0.24}^{+0.47}$ & $-0.31_{-0.15}^{+0.21}$ & $0.22_{-0.31}^{+0.19}$ & $0.0078_{-0.0047}^{+0.0036}$ & $0.52_{-0.21}^{+0.24}$ & $0.31_{-0.27}^{+0.29}$ & $0.61_{-0.44}^{+0.31}$ & $0.30_{-0.39}^{+0.29}$ & $1.43_{-0.13}^{+0.12}$ & $2.34_{-0.34}^{+0.33}$ & $1.355$ \\
\hline
    \end{tabular} 
\end{table*}

\begin{figure*}
\centering{
  \includegraphics[scale=0.8]{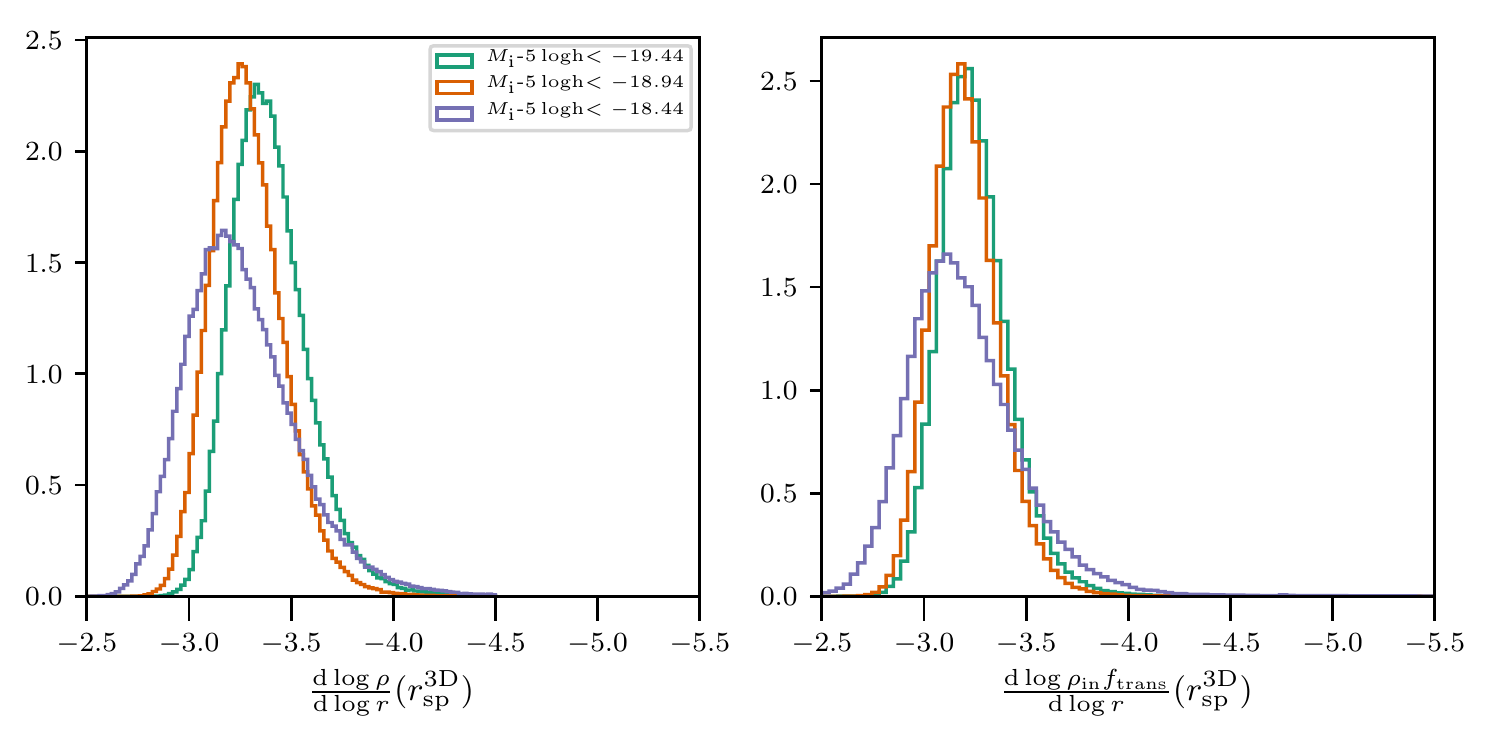}}
\caption{The distributions of the logarithmic derivatives of
the three-dimensional cross-correlation signals at the location of
the splashback radius are shown. In the left panel the distributions of the
derivatives of the full profiles are shown, whereas only the inner
halo term (namely $\rho_{\mathrm{in}}f_{\mathrm{trans}}$) is
shown in the right panel.}
   \label{fig:derivatives} 
\end{figure*}
\begin{figure*}
\centering{
    \includegraphics[scale=0.8]{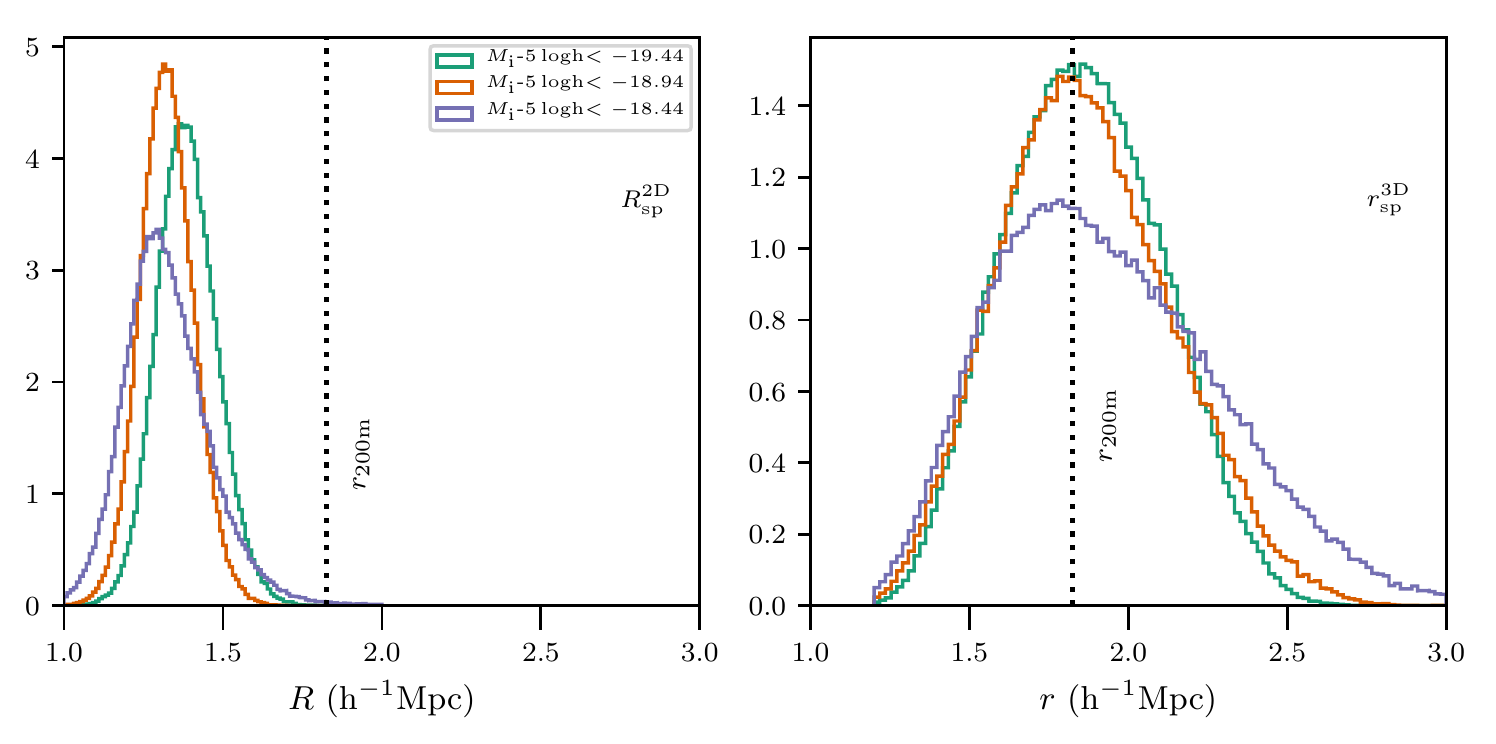}}
\caption{The posterior distributions of the inferred
locations of steepest slope of the two- and three dimensional
cross-correlation signals. The distributions of the
locations of the two-dimensional steepening feature
($R_{\mathrm{sp}}^{\mathrm{2D}}$) are shown in the left panel, while the
distributions of the three-dimensional counterparts
($r_{\mathrm{sp}}^{\mathrm{3D}}$) are shown on the right. The
magnitude limit applied to the galaxy catalogs is indicated for each
distribution. Note that the colors of the distributions match with
the colors used in Figure~\ref{fig:2D_graphs} and
Figure~\ref{fig:3D_graphs}. For comparison, the location of the virial
radius $r_{\mathrm{200m}}$ as calculated from the average cluster
sample properties is indicated by the black, dotted lines.}
   \label{fig:splashback} 
\end{figure*}
\begin{figure}
    \includegraphics[scale=0.65]{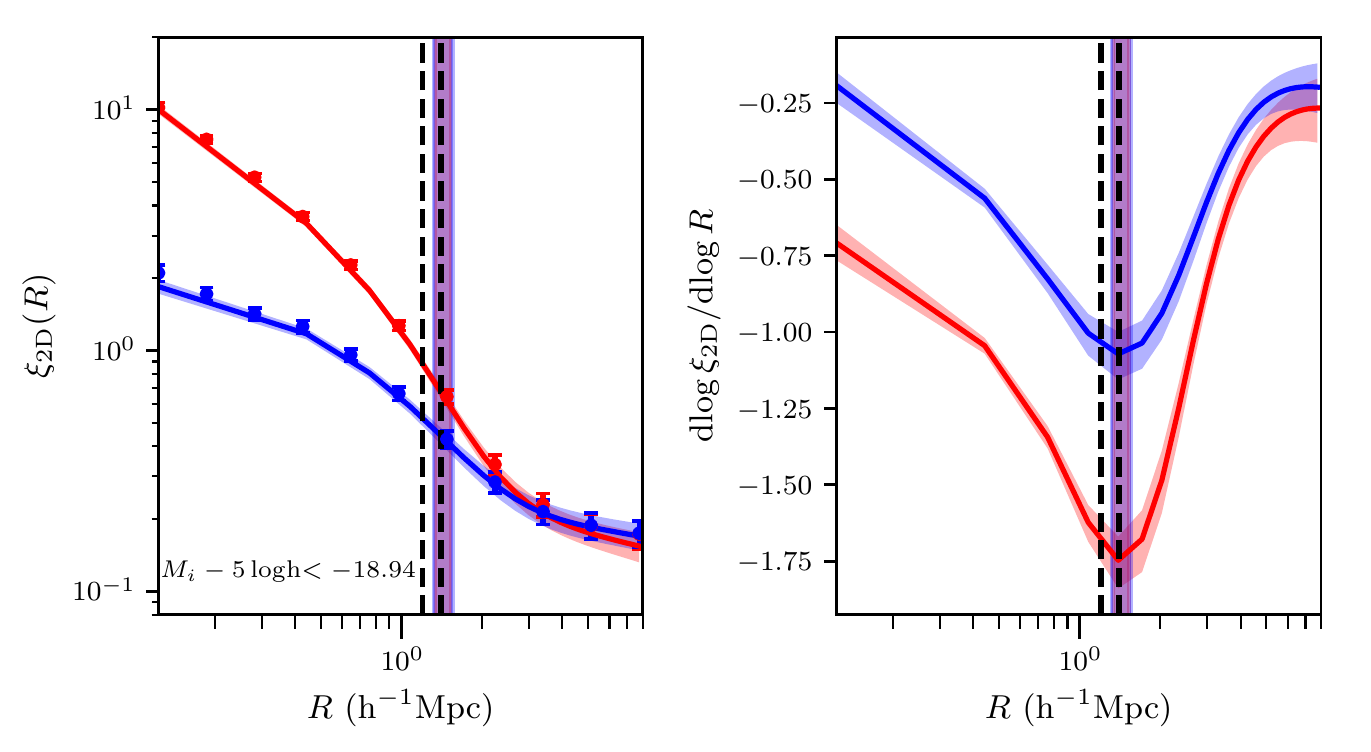}
\caption{The two-dimensional cross-correlations
of the red and blue galaxy populations as inferred by
cross-correlating the cluster sample with the color-separated
subsamples that were extracted from the PS 21.5 galaxy catalog are
shown in the left panel, whereas the associated derivative profiles
are shown on the right. The vertical, shaded bands indicate the
locations of steepest slope of the cross-correlation signal as
inferred from the two subsamples, whereas the black, dashed lines
indicate the upper and lower bounds of the same feature but as
estimated from the full PS 21.5 galaxy catalog. }
   \label{fig:color_curve_2D} 
\end{figure}
\begin{figure*}
    \includegraphics[scale=0.76]{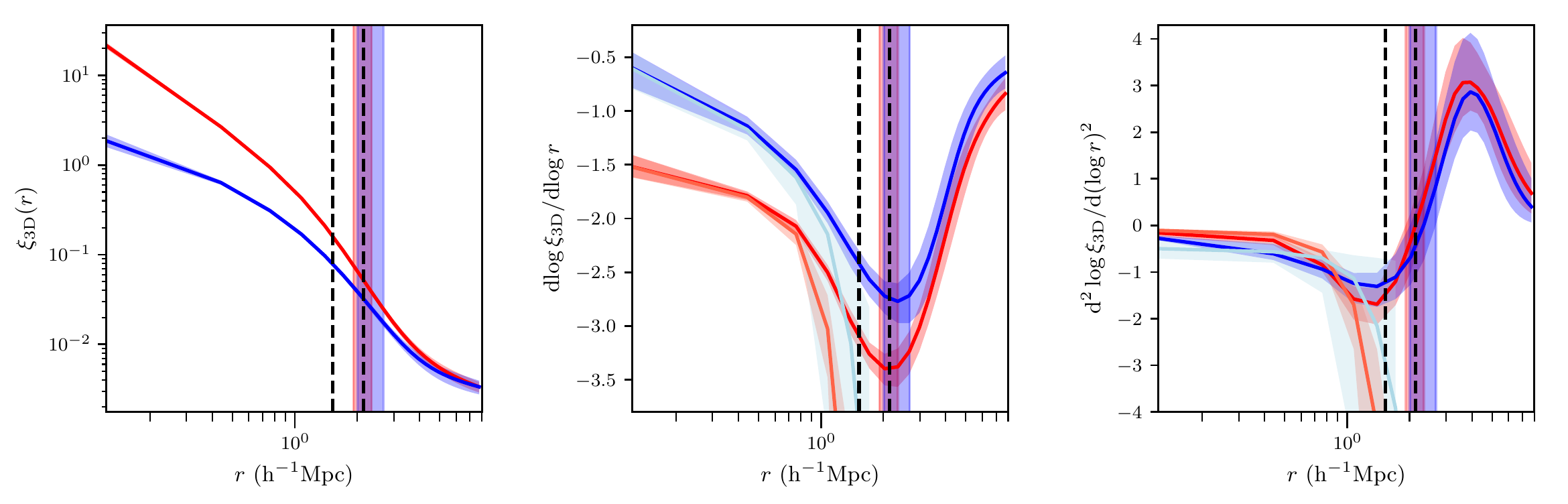}
\caption{We show the three-dimensional cross-correlations as
inferred by cross-correlating the cluster sample with the
color-separated galaxy subsamples, as well as the corresponding
splashback radii (indicated by the colored, vertical bands) and
their first and second order derivative profiles. The light-colored 
curves show the corresponding first and second order derivatives of the one-halo
term (namely $\rho_{\mathrm{in}}(r)f_{\mathrm{trans}}(r)$ in Equation~\ref{eq:dk14}).
Here, the black, dashed lines indicate upper and lower bounds of the 
three-dimensional splashback radius corresponding to the full PS 21.5 galaxy sample.}
   \label{fig:color_curve_3D} 
\end{figure*}
The right panels in Figure~\ref{fig:2D_graphs} show the
corresponding analytical derivatives of the two-dimensional
cross-correlations and the 68\% confidence interval based on our
model fits. The logarithmic derivatives show a distinct steepening
feature at around $1.3~\mpch$. The figure shows that the
location of the steepest slope does not change appreciably when the
magnitude limit is changed, even though we use a sample which is one
magnitude fainter than $M_{\rm i}-5\log h=-19.44$. The 68\%
confidence interval of the location of the steepest slope is
indicated by the vertical, shaded region. We also show the location
of the virial radius $r_{\mathrm{200m}}$ based on the Planck SZ mass estimate 
as a black, dotted line in each panel.

We use the posterior distributions of our model parameters to infer
the three-dimensional cross-correlation and its first and second order logarithmic
derivatives. These inferences along with the corresponding 68\%
confidence intervals are presented in Figure~\ref{fig:3D_graphs},
maintaining the same color scheme as in Figure~\ref{fig:2D_graphs}
for ease of comparison. The three-dimensional cross-correlations
also show significant steepening in each of the cases that we have
explored, reaching logarithmic derivatives steeper than $-3$. The
inferred 68\% confidence regions of the locations of the steepest
slope of the three-dimensional cross-correlations are shown with
vertical, shaded regions in each panel. The posterior
distributions of the locations of the steepest slope in the
two-dimensional and three-dimensional cross-correlations can be
found in the left hand and the right hand panels of
Figure~\ref{fig:splashback}, respectively. The estimates of the splashback radii
for each of the samples are listed in Table~\ref{tab:fit_parameters}
and our results show that the location of the splashback radius does
not depend upon the sample once we use galaxies fainter than $M_{\rm
i}-5\log h=-19.44$. Our measurements have an accuracy of $\sim$15\%.

Following \citet{baxter2017halo}, we also present the values of the
first and second order logarithmic derivatives at the location of the steepest slope for
the total three-dimensional cross-correlations, as well as those for
the inner halo term in the middle and right hand panels of
Figure~\ref{fig:derivatives}, respectively. The logarithmic slope of the 
cross-correlation is significantly steeper than $-3$ at the location 
of the splashback radius, making it difficult to
be reproduced by classical fitting functions like the NFW profile,
which reach such slopes only asymptotically and even that only
without the presence of the outer 2-halo term. This provides
evidence for the existence of the splashback feature. From the second
order derivative profiles we also note that the drop in the cross-correlation
signal is very sharp and happens within a factor 2 in radius.

We perform a preliminary comparison of the measurements with
expectations from cold dark matter models. The average halo mass of
the PSZ2 clusters as estimated from the Sunyaev-Zeldovich signal
is $M_{\rm 500c}=3.0\times10^{14}$ $\msunh$. We convert this mass
estimate to $M_{\rm 200m}=6.2\times10^{14}$ $\msunh$ using the average
concentration mass relation of halos following
\citet{HuKravtsov:2003}. Given the average mass and redshift of our
cluster sample, we calculate the expected splashback radius to be
$1.89$ $\mpch$. We base this estimate on the fitting functions
presented in \citet{more2015splashback}. The splashback radius we
find for the three samples is consistent with this expectation,
although we can not rule out $\sim 15\%$ deviations in either directions,
given our large error bars.

Next, we present our measurements of the projected and
three-dimensional cross-correlations of the red and blue galaxy populations
satisfying $M_{i}-5\log h<-18.94$ with our SZ-selected cluster
sample. The projected cross-correlations are presented in Figure~\ref{fig:color_curve_2D},
whereas the three-dimensional counterparts are shown in 
Figure~\ref{fig:color_curve_3D}. The shaded regions show
the 68\% confidence intervals from our fits. The vertical, shaded
bands with different colors indicate the 68\% confidence regions of
the locations of the steepest slope of the two- and the
three-dimensional cross-correlations. The black dashed lines show
these ranges for the entire galaxy sample without regard to color.
The best fit parameters as well as the inferred splashback radii are
listed in Table~\ref{tab:fit_parameters}. We
see that the red galaxies have a steeper cross-correlation profile
than the blue galaxies. Although there seems to be  a tendency for the red
galaxies to have a smaller splashback radius, the differences we see
are not statistically significant given the current errors. The
slopes of the three-dimensional cross-correlations reach values
steeper than -2.5 at the splashback radius, which is difficult to model with
traditional NFW profiles. We take the detection of those steepening features 
as evidence for the existence of the splashback feature in both galaxy populations.
Taken at face value, our finding of the splashback feature in the blue
population would imply that there needs to be a reasonable fraction of
blue galaxies that fall into the cluster and continue to stay blue
even after reaching their apocenters.

We caution however that the cut we use to define our samples was
defined based on spectroscopic galaxies, which tend to be brighter
than the typical galaxies we use for the cross-correlation. As we use
galaxies at fainter magnitudes for the cross-correlation, the
photometric errors increase dramatically and can scatter many more of
the red galaxies into the blue galaxy sample. We report on
our investigations in Appendix~\ref{sec:Errors} as a cautionary tale,
and to guide future efforts to establish the splashback radius in the
blue galaxy population. We show that given the current state of the
data we can marginally exclude a contamination of the correlation
function of blue galaxies from the red galaxy population but
appropriate caution is warranted in the interpretation or use of the
results derived using the color separated galaxy populations.

\section{Conclusions and future directions}
\label{sec:Conclusions}
The splashback radius of dark matter halos is a unique observational
probe of the mass accretion rate of dark matter halos. Although the
splashback radius has been well-characterized in simulations, the
observational evidence for the splashback radius presented using
optical cluster catalogs has come under intense scrutiny. In this
work, we tackle this issue by searching for evidence of the
splashback radius in galaxy clusters found by the Planck surveyor using
the thermal Sunyaev-Zeldovich effect. The use of this sample avoids
the circularity of using photometric galaxy catalogs to identify
clusters as well as to detect the splashback radius. 

We cross-correlate these clusters with photometric galaxies from the
Pan-STARRS survey to obtain the two-dimensional cross-correlation
function and search for evidence for the splashback feature.
Additionally, we divided our galaxy catalog into two subsamples of
red and blue galaxies and investigated the cross-correlations of the
two subsamples with the clusters, separately.

Our main findings can be summarized as follows:
\begin{itemize}
\item We detect a clear signature of a steepening feature in the
cross-correlation of Planck SZ clusters with Pan-STARRS
galaxies. The steepest logarithmic slopes that we find in our cross-correlation
signals are steeper than $-3$, and would hence be poorly fit by the
NFW profile. We associate this steepening with the splashback feature.
\item The location of the inferred splashback radius is $r_{\rm
sp}=1.85_{-0.30}^{+0.26}$ $\mpch$, which is consistent with expectations
from numerical simulations for
halos of an average mass $M_{\rm 500c}=3.0\times10^{14}\msunh$ at an
average redshift of $z=0.18$ in a collision-less dark matter Universe. 
However, given the errors we cannot
currently rule out $\sim15\%$ deviations from these expectations.
\item We find that the location of the steepest slope does not
strongly depend on the magnitude of the galaxy samples we use, once
we go fainter than $M_i-5\log h=-19.44$.
\item By separately studying the cross-correlation of red and blue
galaxies with the clusters, we present evidence for the presence of
the splashback feature in both populations. The existence of the
splashback radius for the 
star forming galaxy population could be of significance for the
models of satellite quenching in galaxy clusters. However, photometric
errors hinder a clean interpretation of the signal.
\end{itemize}

The signal-to-noise ratio of the current measurement is a result of
the limited depth of the Pan-STARRS catalog, as well as the limited
number of galaxy clusters detected using the SZ effect. As we consider
fainter galaxy catalogs, the sky fraction in which the Pan-STARRS
galaxy catalogs are complete reduces. Furthermore, the contamination
from background galaxies is expected to increase at deeper magnitudes
as well. A more precise estimation of the
uncorrelated component would be possible by using a random galaxy
catalog in addition to the random cluster sample. The masking
information in Pan-STARRS is not currently easily accessible, which
prevents the use of the more sophisticated Landy \& Szalay estimator. 
Increasing the redshift range of clusters could potentially yield a
bigger cluster sample, but would require us to use a sample of
galaxies with a brighter absolute magnitude limit for galaxies, 
which reduces the number of galaxies that can be used to infer the
cross-correlation signal. We are exploring the use of alternative
cluster catalogs such as those detected from Xray surveys.

The separation of the blue and red galaxy populations as described in
Section~\ref{sec:Color} is prone to photometric errors as explored in
the Appendix~\ref{sec:Errors}. A deeper galaxy catalog would be required in order to
confirm or rule out the existence of the splashback feature in the
blue galaxy population. The ongoing deep galaxy surveys such as the
Hyper Suprime-Cam \citep{Aihara2018} and the Dark Energy Survey \citep{Abbott2018} would be able
to provide such data sets, albeit in a limited area. The Large
Synoptic Survey telescope \citep{LSST2009} would eventually provide deep as well
as wide galaxy catalogs to eventually establish the locations of the
splashback radius at high significance.

Lastly, but most importantly, the investigation of any systematics
which might originate from the SZ selection of the Planck clusters
is beyond the scope of the current work. We caution that there may
be residual systematics in the selection which could affect the
interpretation of our measurements. We plan to investigate such
selection systematics with the help of hydrodynamical simulations.

Our curated galaxy catalogs from the Pan-STARRS survey for different
depths and the corresponding masks are available upon request.

While this work was in preparation, we became aware of a related study
by Shin et al. (2018). Our results are complementary to theirs given
the different data samples and cluster catalogs.

\section*{Acknowledgements}
DZ is grateful to Kavli IPMU for its hospitality during this research
project. SM is supported by a grant-in-aid by the Japan Society for
Promotion of Science (JSPS), grant number 16H01089. We acknowledge
useful discussions regarding this work with Benedikt Diemer, Neal
Dalal, Bhuvnesh Jain, Andrey Kravtsov, as well as participants of the
KITP UCSB workshop on the Galaxy-Halo connection in 2017. SM was
supported in part by the National Science Foundation under Grant No.
NSF PHY-1125915 during this workshop.

The Pan-STARRS1 Surveys (PS1) and the PS1 public science archive
have been made possible through contributions by the Institute for
Astronomy, the University of Hawaii, the Pan-STARRS Project Office,
the Max-Planck Society and its participating institutes, the Max
Planck Institute for Astronomy, Heidelberg and the Max Planck
Institute for Extraterrestrial Physics, Garching, The Johns Hopkins
University, Durham University, the University of Edinburgh, the
Queen's University Belfast, the Harvard-Smithsonian Center for
Astrophysics, the Las Cumbres Observatory Global Telescope Network
Incorporated, the National Central University of Taiwan, the Space
Telescope Science Institute, the National Aeronautics and Space
Administration under Grant No. NNX08AR22G issued through the
Planetary Science Division of the NASA Science Mission Directorate,
the National Science Foundation Grant No. AST-1238877, the
University of Maryland, Eotvos Lorand University (ELTE), the Los
Alamos National Laboratory, and the Gordon and Betty Moore
Foundation.

Based on observations obtained with Planck
(http://www.esa.int/Planck), an ESA science mission with instruments
and contributions directly funded by ESA Member States, NASA, and
Canada.  Some of the results in this paper have been derived using
the HEALPix (K.M. Górski et al., 2005, ApJ, 622, p759) package

\bibliographystyle{mnras}
\bibliography{RSP_library} 

\appendix

\section{Sky maps}
\label{sec:figures}
\begin{figure*}
    \includegraphics[width= \textwidth]{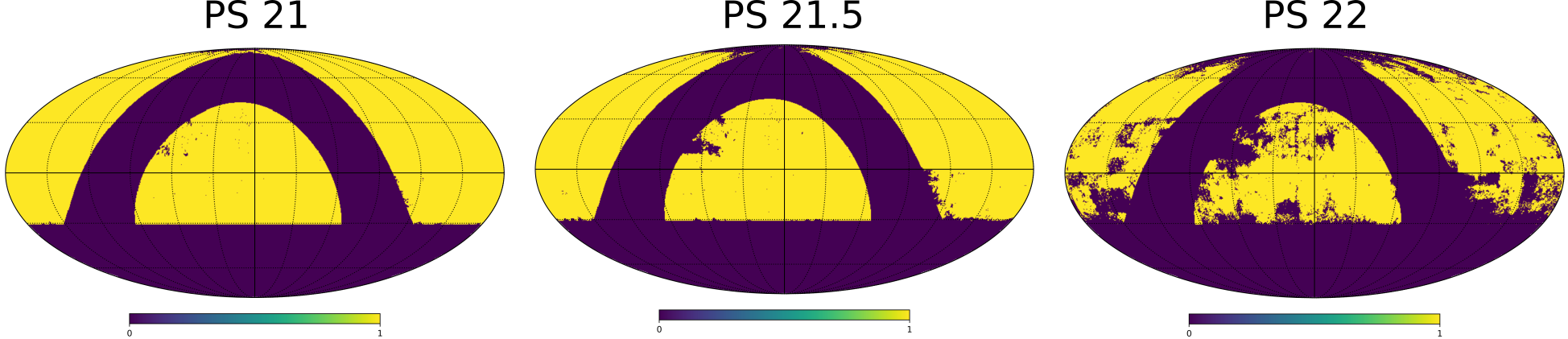}
\caption{Sky maps displaying the masked out regions of the used
galaxy catalogs extracted from the Pan-STARRS 3$\pi$ Steradian
survey in purple. The ICRS coordinate system is used.}
   \label{fig:heal_map} 
\end{figure*}

We present the sky locations of the SZ selected clusters from the PSZ2
catalog that we used in our study. The purple areas mark the masked
out regions as defined by the union selection function of the PSZ2
catalog. The sky maps in Figure~\ref{fig:heal_map} show the survey
masks of our three galaxy catalogs.  All sky maps presented in this
section use the ICRS coordinate system.

\begin{figure}
\centering{
    \includegraphics[scale=0.8]{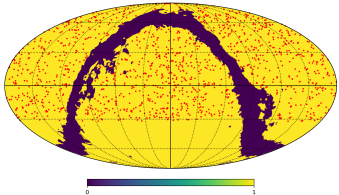}}
\caption{Sky map showing the sky positions of the used clusters as
selected from the PSZ2 catalog. In total 596 clusters have been
selected. The purple areas mark the regions that are excluded by the
union survey selection function. Note that all clusters with
$\delta<-31\degr$ have been removed since this region is not covered
by the Pan-STARRS 3$\pi$ Steradian survey. The ICRS coordinate
system is used.}
    \label{fig:planck_fig} 
\end{figure}

\section{Modelling cluster mis-centering}
\label{sec:mis-centering}
As can be seen from the analysis and results in \citet{baxter2017halo},
the mis-centering of the central cluster positions in optical clusters
were not large enough to change the location of the steepest slope in
the cluster-galaxy cross-correlations. They do however decrease the
significance of the finding. Therefore, we explore the effects of
mis-centering on the correlation function and its influence on the
results. 

The mass dependence of the exact mis-centering fractions (for the
brightest galaxy to be the central) are not well understood and
mis-centering effects are therefore difficult to model
\citep{Skibba:2011, Hoshino:2015}. We follow the approach outlined in
\citet{baxter2017halo} to take the effects of a possible mis-centering
of a fraction of the cluster positions into account by modelling the
influence on the two-dimensional correlation function.

If a fraction $f_{\rm mis}$ of the galaxy clusters in our sample are
mis-centered, then the measured correlation function $\xi'_{\rm
2D}(R)$ is given by
%
%
\begin{equation} 
\xi'_{\rm 2D}(R) = (1-f_{\rm mis})\xi_{\rm 2D}(R)+f_{\rm mis}\xi_{\rm
(2D,mis)}(R)\\,
\label{eq:full}
\end{equation}
where $\xi_{\rm (2D,mis)}$ denotes the contribution of the
mis-centered clusters to the correlation function, and $\xi_{\rm
2D}(R)$ corresponds to the contribution of the correctly centered
clusters. We model the mis-centered component $\xi_{\rm (2D,mis)}$ as
\begin{equation}
\xi_{\rm (2D,mis)}(R)=\int_0^\infty \mathrm{d}R_{\rm mis} P(R_{\rm
mis})\xi_{\rm (2D,mis)}(R|R_{\rm mis})\,,
\end{equation}
where $P(R_{\rm mis})$ denotes the probability that a cluster is
centered at a comoving distance $R_{\rm mis}$ from the brightest
galaxy. The contribution $\xi_{\rm (2D,mis)}(R|R_{\rm mis})$ is
related to the correlation function of the correctly centered clusters
$\xi_{\rm 2D}$ as
\begin{equation}
\xi_{\rm (2D,mis)}(R|R_{\rm
mis})=\int_0^{2\pi}\frac{\mathrm{d}\theta}{2\pi}\xi_{\rm
2D}\left(\sqrt{R^2+R_{\rm mis}^2+2RR_{\rm mis}\cos{\theta}}\right)\,,
\end{equation}
according to \citet{yang2006weak} and \citet{johnston2007cross}. We
model the mis-centering probability $P(R_{\rm mis})$ as a Rayleigh
distribution,
\begin{equation}
P(R_{\rm mis})=\frac{R_{\rm mis}}{\sigma^2}\exp{\left( -\frac{R_{\rm
mis}^2}{2\sigma^2}\right)}\,,
\end{equation}
Thus, the mis-centered contribution is fully characterized by the
two parameters $f_{\rm mis}$ and the witdh of the mis-centering
probability distribution $\sigma$.

To find the priors for the two mis-centering paramters $f_{\rm mis}$
and $\sigma$ we cross-match our SZ selected cluster sample with the
X-Ray selected ACCEPT cluster sample \citep{cavagnolo2009vizier}.
Assuming the ACCEPT clusters to lie at the minimum of the
gravitational potential, we infer Gaussian priors for the two
mis-centering model parameters $f_{\rm mis}=0.15\pm0.21$ and
$\sigma=0.41\pm0.30$.

To study the influence of mis-centering on our findings we repeat the
MCMC model fitting using our new model including mis-centering given
in Equation~\ref{eq:full}, which now includes two more model
parameters. 
\begin{table}
    \centering
    \caption{Listing of the splashback radii inferred by including the effects of mis-centering of the central cluster positions. }
    \label{tab:mis_splash}
    \begin{tabular}{ccc}
    \hline 
gal cat &  $R_{\mathrm{sp}}^{\mathrm{2D}}$ & $r_{\mathrm{sp}}^{\mathrm{3D}}$ \\
\hline 
\hline 
PS 21 &$1.35_{-0.12}^{+0.13}$ & $1.76_{-0.33}^{+0.29}$\\
\hline
PS 21.5 & $1.30_{-0.10}^{+0.10}$&$1.81_{-0.33}^{+0.28}$\\
\hline
PS 22 &$1.33_{-0.17}^{+0.13}$ & $1.96_{-0.45}^{+0.37}$\\
\hline
PS 21.5 (R) & $1.45_{-0.11}^{+0.12}$ & $2.23_{-0.26}^{+0.27}$ \\
\hline
PS 21.5 (B) & $1.42_{-0.25}^{+0.20}$ & $2.39_{-0.39}^{+0.39}$ \\
\hline
    \end{tabular} 
\end{table}

We find that the use of a projected mis-centering model increases the
errorbars on the predicted confidence intervals for the projected
cross-correlation signal by $\sim 25$\%. However, the effect is barely
noticeable in the three-dimensional profiles.  The inclusion of
mis-centering is reflected as a small increase in the inferred errors
of the projected and three-dimensional splashback radii as seen in
Table~\ref{tab:mis_splash}. Although we notice shifts in the inferred
central values of the splashback radii, none of these shifts appear
systematic or significant given the errors.

\section{Cornerplots}
\label{sec:cornerplots}
\begin{figure*}
    \includegraphics[width= \textwidth]{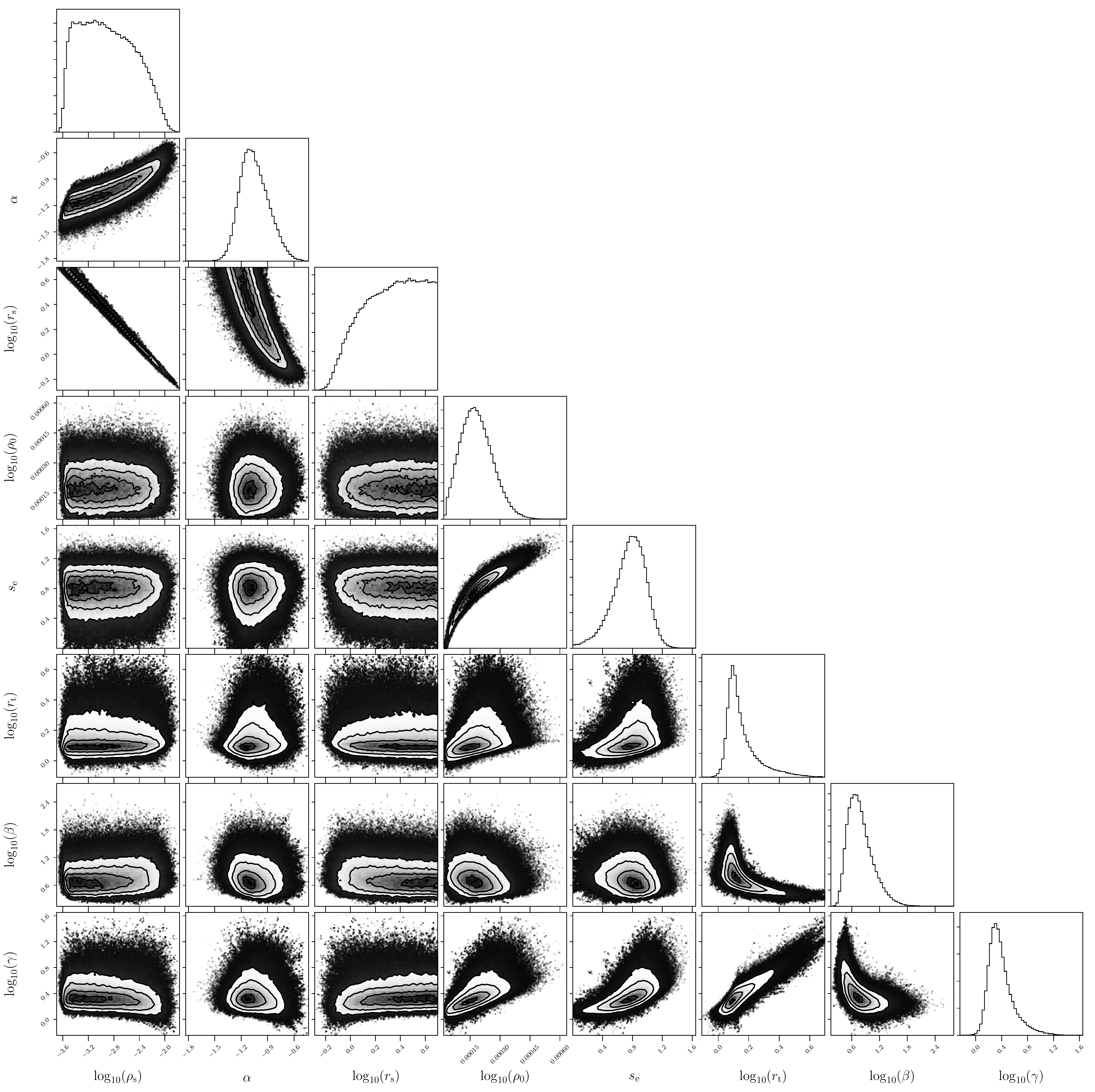}
\caption{The two-dimensional posterior distributions of each pair of
the fitting parameters corresponding to the functional
form in Equation~\ref{eq:full} for the PS 21 sample. 
}
   \label{fig:corner_21} 
\end{figure*}

\begin{figure*}
    \includegraphics[width= \textwidth]{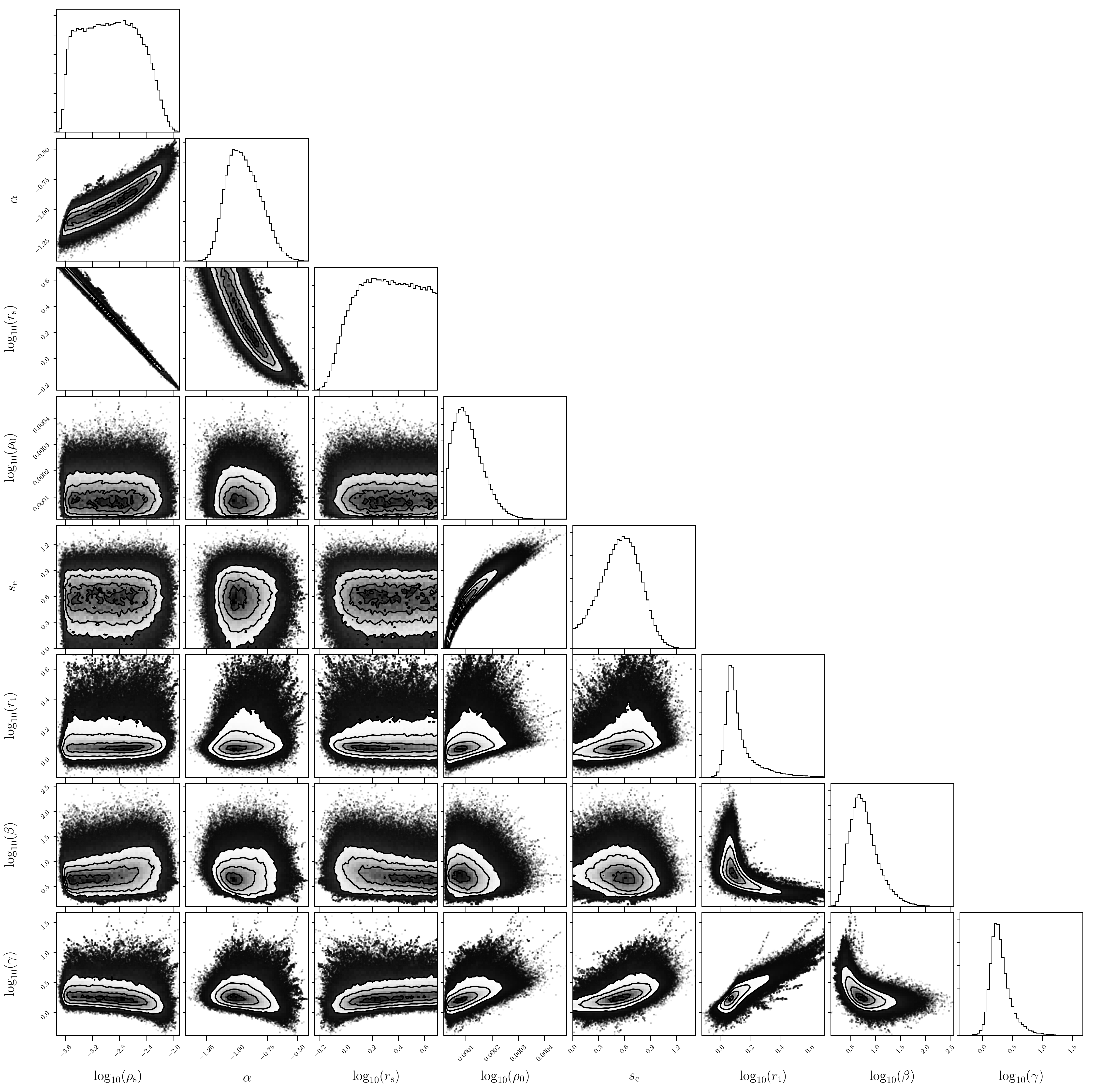}
\caption{The two-dimensional posterior distributions of each pair of
the fitting parameters corresponding to the functional
form in Equation~\ref{eq:full} for the PS 21.5 sample.
}
   \label{fig:corner_21.5} 
\end{figure*}

\begin{figure*}
    \includegraphics[width= \textwidth]{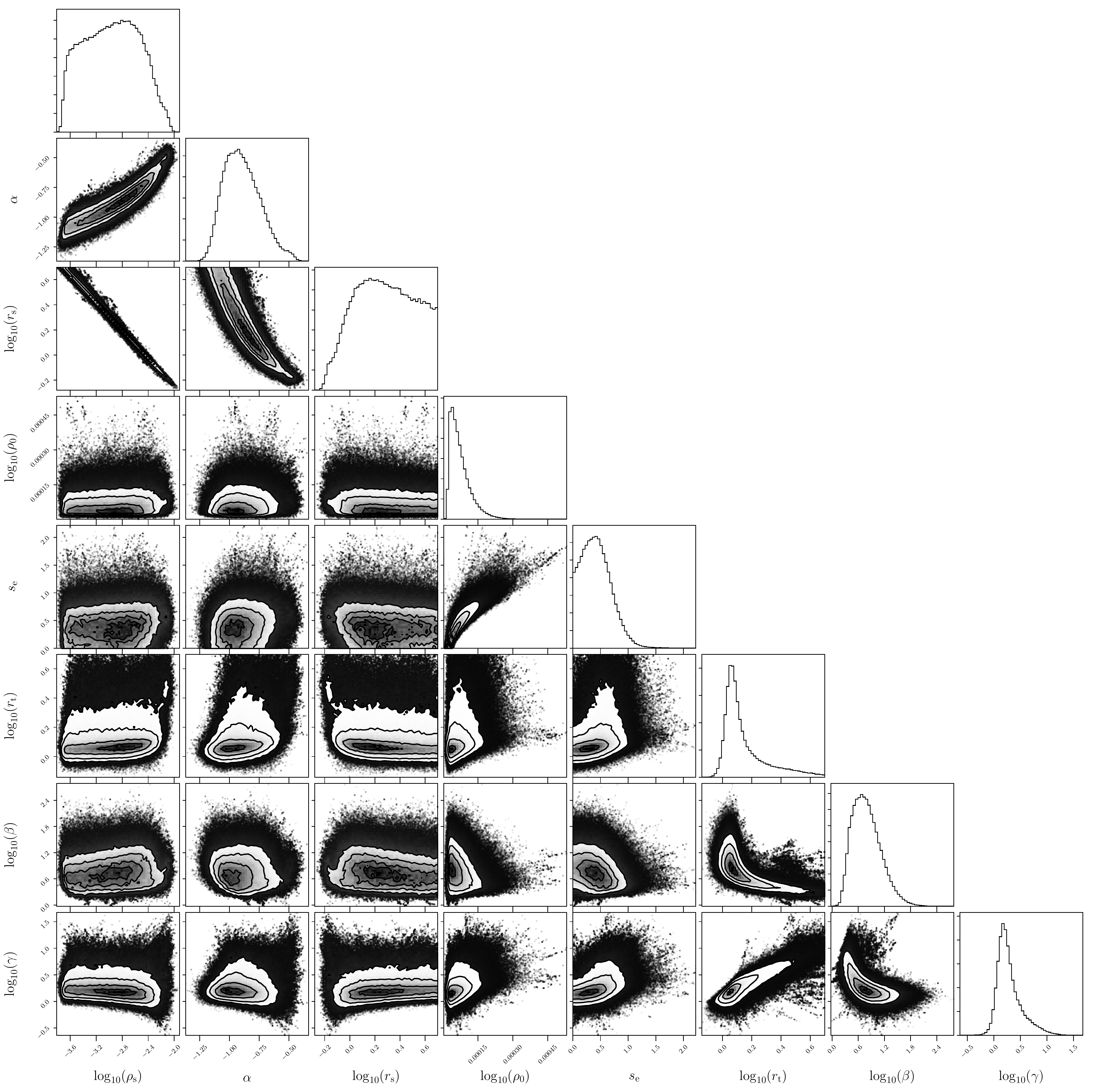}
\caption{The two-dimensional posterior distributions of each pair of
the fitting parameters corresponding to the functional form in
Equation~\ref{eq:full} for the PS 22 sample.}
   \label{fig:corner_22} 
\end{figure*}

\begin{figure*}
    \includegraphics[width= \textwidth]{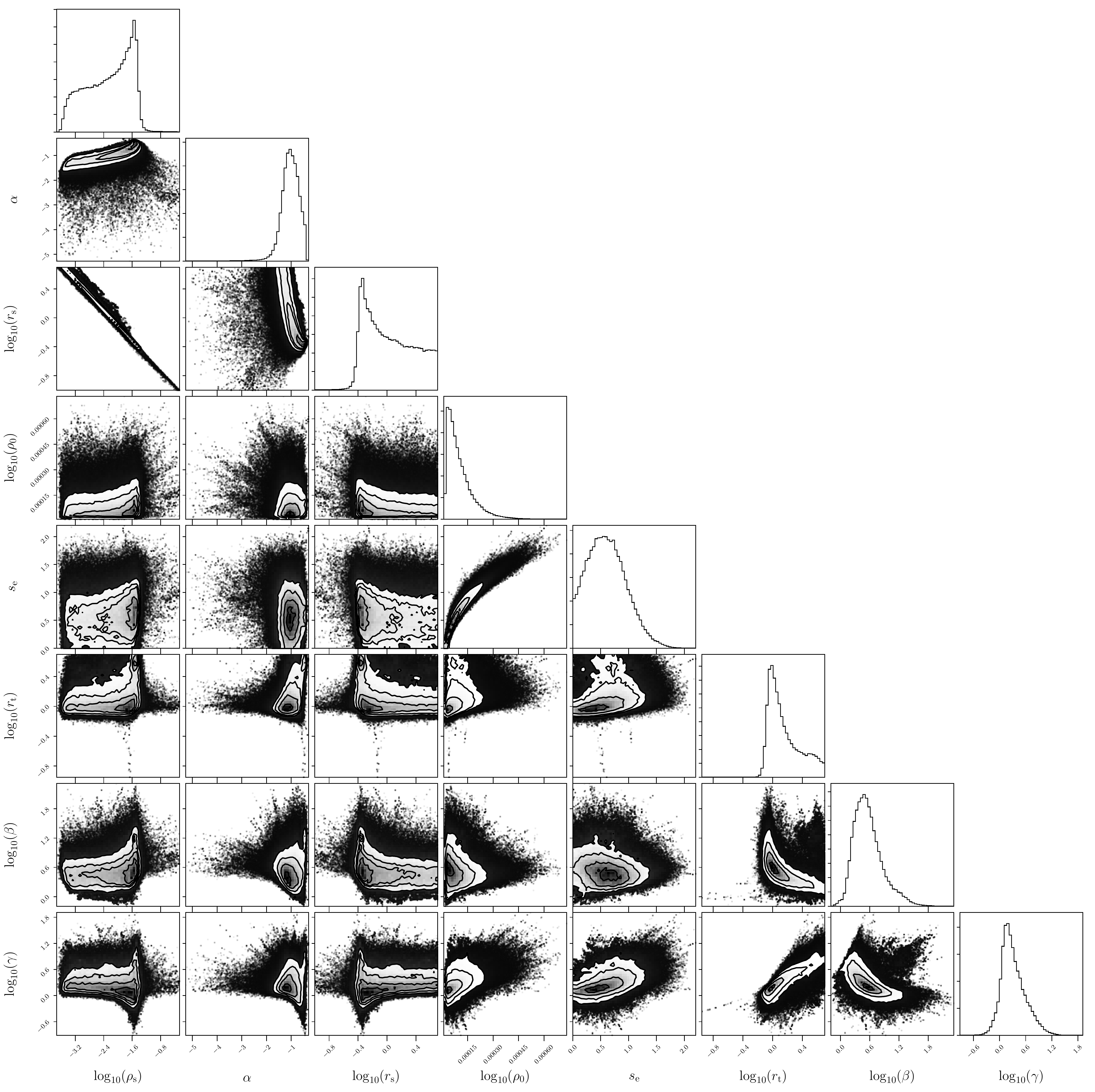}
\caption{The two-dimensional posterior distributions of each pair of
the fitting parameters corresponding to the functional form in
Equation~\ref{eq:full} for the PS 21 sample, where we restricted our
cross-correlatoin analysis to the red galaxy population.
}
   \label{fig:cornerred} 
\end{figure*}

\begin{figure*}
    \includegraphics[width= \textwidth]{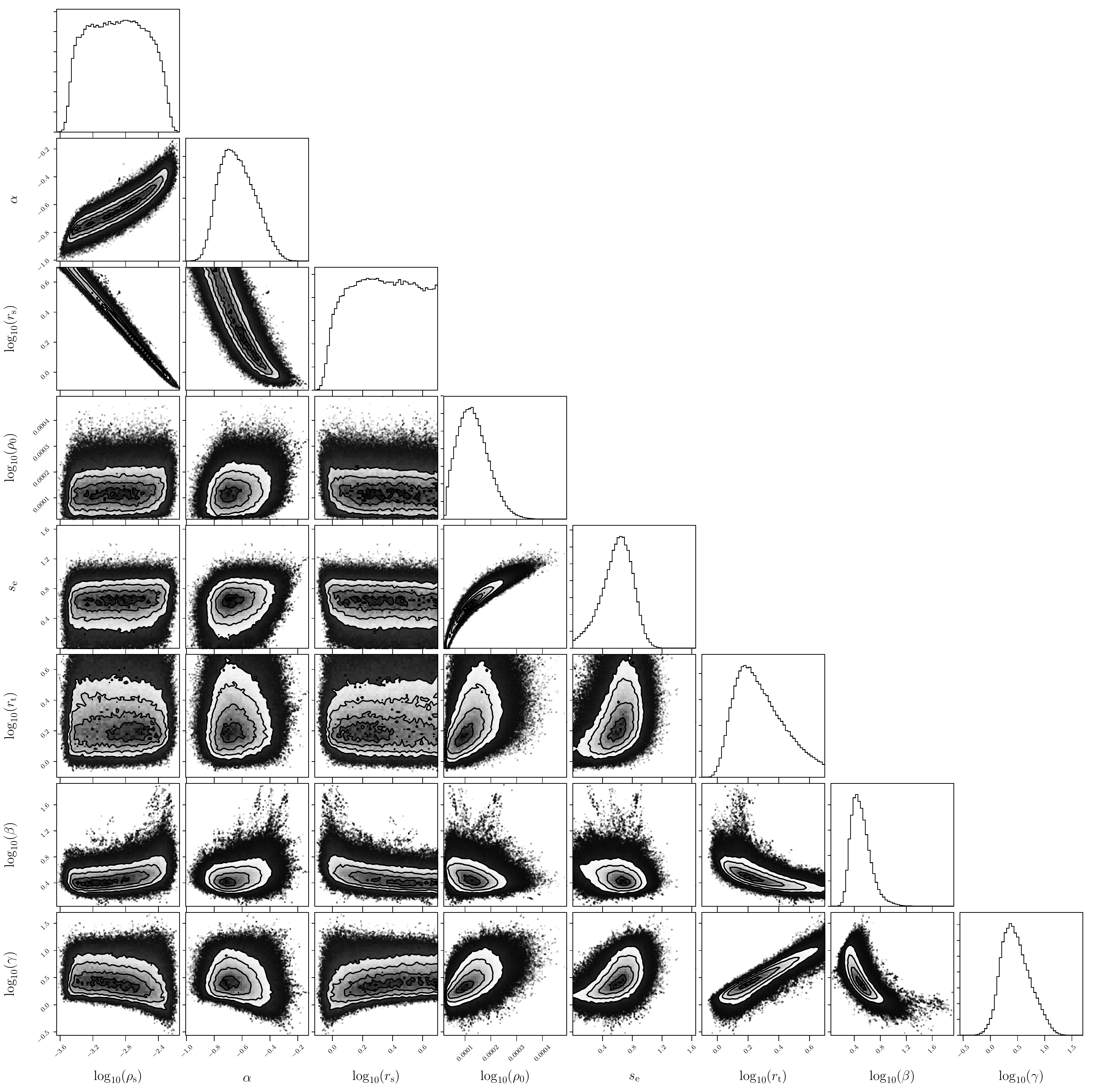}
\caption{The two-dimensional posterior distributions of each pair of
the fitting parameters corresponding to the functional form in
Equation~\ref{eq:full} for the PS 21 sample, where we restricted our
cross-correlatoin analysis to the blue galaxy population. 
}
   \label{fig:cornerblue} 
\end{figure*}

In Figures \ref{fig:corner_21} to \ref{fig:cornerblue} we present
the two-dimensional posterior distributions for each pair of the
fitting parameters corresponding to the functional
form in Equation~\ref{eq:full}. We obtained the distributions by  using the
affine invariant Markov Chain Monte Carlo sampler of
\citet{goodman2010ensemble} as implemented in the parallel python
package {\it emcee} by \citet{foreman2013emcee}.

\section{Estimating the contamination of the blue galaxy population}
\label{sec:Errors}

\begin{figure*}
\centering{
    \includegraphics[width= 0.5\textwidth]{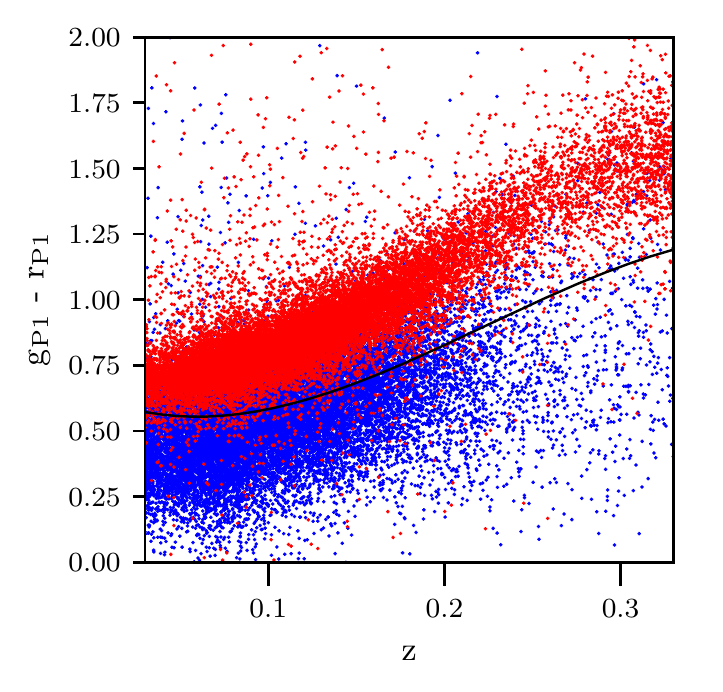}}
\caption{Scatter plot showing the $(g_{\mathrm{P1}}-r_{\mathrm{P1}})$ colors
of the matched galaxies versus their redshifts, where the red dots correspond
to galaxies identified as red from the SDSS spectroscopic colors and blue dots
correspond to galaxies identified as blue, respectively. The black, solid line
indicates the spline cut separating the two populations and excluding the red 
galaxies from the blue population at a confidence level of $3 \sigma$.}
   \label{fig:g-r_vs_z} 
\end{figure*}

To infer the color cut which separates star forming (blue) galaxies from the
quenched (red) galaxy population, we cross-matched the SDSS spectroscopic
sample to its Pan-STARRS photometry. In Figure~\ref{fig:g-r_vs_z}, we show the
plot of the Pan-STARRS $(g_{\mathrm{P1}}-r_{\mathrm{P1}})$ color as a 
function of redshift, and the color cut we use in our analysis. 
The SDSS spectroscopic sample is however quite shallow, 
and thus consists of galaxies that are brighter than the Pan-STARRS 
galaxies that we wish to cross-correlate. At fainter magnitudes, we can 
expect red galaxies to scatter into the blue population 
due to photometric errors, potentially contaminating the
correlation function measurement. This could erroneously cause the
splashback feature observed in the blue population. Therefore, we
need to assess the possiblity of such a contamination. 

\begin{figure*}
\centering{
    \includegraphics[width= \textwidth]{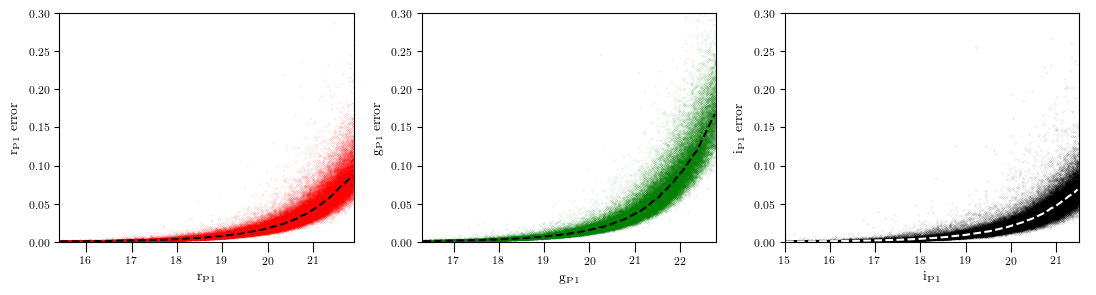}
\caption{Shown are the photometric errors of the Pan-STARRS galaxies versus
their magnitude in each of the used band. The dashed curves indicate the running
median of the photometric error.}
   \label{fig:mag_errors} 
}
\end{figure*}

In Figure~\ref{fig:mag_errors}, we show the photometric errors of the Pan-STARRS
galaxies as a function of magnitude in each of the bands we use. The
galaxies that we use in our cross-correlation analysis for the red/blue
galaxy population have an upper absolute magnitude limit of $M_{i}-5\log h=-18.94$. This
corresponds to a different apparent magnitude limit at each redshift.
For each spectroscopically matched galaxy, we figure out how faint the
apparent magnitude limit in the $i_{\mathrm{P1}}$ band is compared to the actual
apparent magnitude of the galaxy, recorded by Pan-STARRS. Assuming that the populations of
galaxies that we use to cross-correlate with clusters have the same
intrinsic colors, but are just fainter, we infer the true
intrinsic magnitude of the galaxies that we cross-correlate in the
$g_{\mathrm{P1}}$ and the $r_{\mathrm{P1}}$ band. We randomly perturb these magnitudes 
by the expected photometric errors at that magnitude as reported in Figure~\ref{fig:mag_errors}.

Next, we compute the fraction of galaxies which were intrinsically red
that now entered the blue population by the perturbation, where we
defined the separation of the populations using the same color
cut. We find such a contamination of intrinsically red galaxies to
the blue population to be about 5\%. This value would apply for a galaxy population
measured in the field. Given that the red fraction is higher in clusters ($\sim
60$\% in the whole cluster) than in the field ($\sim 40$\%),
we expect there to be a larger proportion of red galaxies in the cluster which could
potentially contaminate the blue galaxies due to photometric errors.
This would roughly double the contamination to be about 10\%.

This is a conservative estimate of the contamination, because the
spectroscopic galaxy sample we use has a large incompleteness at the
blue end at redshifts beyond $0.2$, where SDSS targetted the luminous
red galaxy population, recording only very few blue galaxies. 
Thus we are missing a lot of the blue galaxy
population in the highest redshift bins in our spectroscopic sample.
Nevertheless we assess if a $\sim 10$\% contamination could
cause the splashback signal found in the cross-correlation for the blue galaxies.


The null hypothesis that we would like to establish or rule out is
that the blue galaxies just consist of the infall population and show
no splashback feature. We assume a simple $r^{-1.5}$ power law for the
3D cross-correlation of the blue population consistent for clusters with
an infalling population, and a 3D cross-correlation equal to the signal
found for the red galaxy population in our analysis. Given these
cross-correlation functions we can estimate the cross-correlation of
the blue population when contaminated with a given amount of red
galaxies, by considering a weighted sum of the two signals. 
We find that in order to reproduce the observed
cross-correlation function of the blue galaxies, we need a contamination
of about 20\%. While our conservatively estimated contamination is
smaller than this value, a proper modeling of this contamination is
warranted before definitive conclusions can be drawn. Larger galaxy
surveys with better photometry such as the HSC or LSST would be of
significant help in alleviating these issues.


\label{lastpage}
\end{document}